\documentclass[journal=jpclcd]{achemso}%
\usepackage{dcolumn}
\usepackage{bm}
\usepackage[version=4]{mhchem}
\usepackage{siunitx}
\usepackage{subfigure}
\usepackage{amsfonts,amsmath,amssymb, subfigure}
\usepackage{color}
\usepackage{commath}
\usepackage{esint}
\usepackage[T1]{fontenc}
\usepackage{graphicx}
\usepackage[utf8]{inputenc}
\usepackage{microtype}
\usepackage{times}
\usepackage{hypernat}
\usepackage[colorlinks = true, linkcolor = blue, urlcolor  = blue, citecolor = blue, anchorcolor = blue]{hyperref}
\usepackage[textsize=footnotesize,obeyFinal]{todonotes}
\usepackage{xspace}
\usepackage{textcomp}
\usepackage{diagbox}
\usepackage{lineno}

\newcommand*{\bra}[1]{\ensuremath{\left<#1\right|}\xspace}%
\newcommand*{\ket}[1]{\ensuremath{\left|#1\right>}\xspace}%

\usepackage{xspace}

\usepackage[utf8]{inputenc}
\usepackage[T1]{fontenc}
\usepackage{mathptmx}
\usepackage{etoolbox}

\usepackage{color}
\definecolor{oldtxtcolor}{rgb}{0.00, 0.0, 0.5}
\definecolor{newtxtcolor}{rgb}{0.00, 0.3867, 0.00}
\definecolor{newtxtcolor}{rgb}{0.00, 0.0, 1.0}
\definecolor{oldtxtcolor}{rgb}{1.00, 0.0, 0.00}

\def\verX{12}
\def\verO{1}
\def\verN{2}
\def\verON{12}

\ifx\verX\verO
\newcommand { \oldtxt }[1] {{\color{oldtxtcolor}{#1}}}
\newcommand { \newtxt }[1] {}
\fi
\ifx\verX\verN
\newcommand { \oldtxt }[1] {}
\newcommand { \newtxt }[1] {{\color{newtxtcolor}{#1}}}
\fi
\ifx\verX\verON
\newcommand { \oldtxt }[1] {{\color{oldtxtcolor}{#1}}}
\newcommand { \newtxt }[1] {{\color{newtxtcolor}{#1}}}
\fi

\title{Ultrafast X-ray Diffraction Probe of Coherent Spin-state Dynamics in Molecules}

\author{Xiaoyu Mi}
\affiliation{State Key Laboratory for Mesoscopic Physics and Collaborative Innovation Center of Quantum Matter, School of Physics, Peking University, Beijing 100871, China}
\author{Ming Zhang}
\affiliation{State Key Laboratory for Mesoscopic Physics and Collaborative Innovation Center of Quantum Matter, School of Physics, Peking University, Beijing 100871, China}
\author{Zheng Li}
\email{zheng.li@pku.edu.cn}
\affiliation{State Key Laboratory for Mesoscopic Physics and Collaborative Innovation Center of Quantum Matter, School of Physics, Peking University, Beijing 100871, China}
\alsoaffiliation{Collaborative Innovation Center of Extreme Optics, Shanxi University, Taiyuan, Shanxi 030006, China}
\altaffiliation{Peking University Yangtze Delta Institute of Optoelectronics, Nantong, Jiangsu 226000, China}
\begin{document}
	\begin{tocentry}
		\begin{center}
			\includegraphics[width=2in]{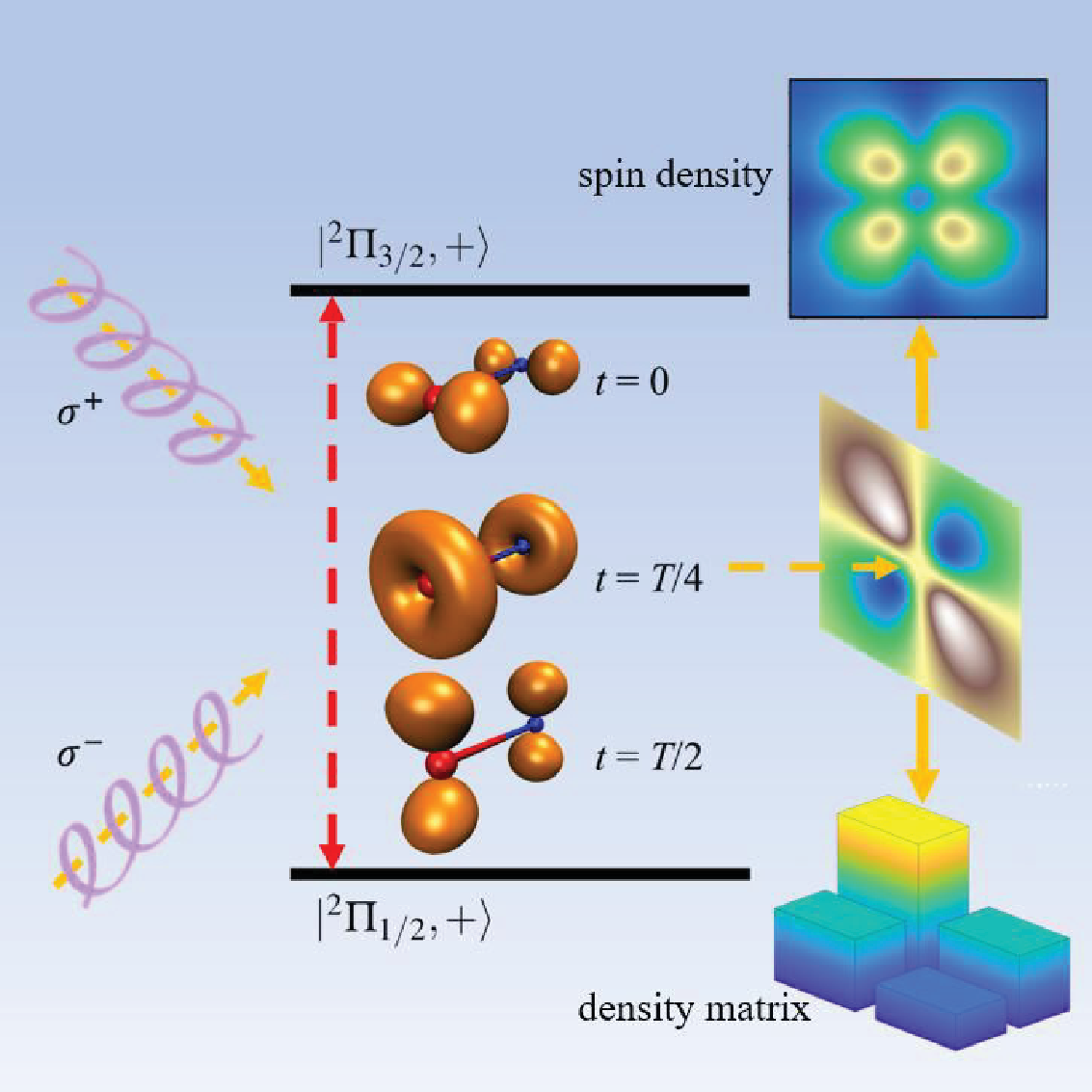}
		\end{center}
	\end{tocentry}
	\begin{abstract}
		We propose an approach to probe coherent spin-state dynamics of molecules using circularly polarized hard x-ray pulses. 
		For the dynamically aligned nitric oxide molecules in a coherent superposition spin-orbit coupled electronic state that can be prepared through stimulated Raman scattering, we demonstrate the capability of ultrafast x-ray diffraction to not only reveal the quantum beating of the coherent spin-state wave packet, but also image the spatial spin density of the molecule.
		With circularly polarized ultrafast x-ray diffraction signal, we show that the electronic density matrix can be retrieved.
		The spatio-temporal resolving power of ultrafast x-ray diffraction paves the way for tracking transient spatial wave function in molecular dynamics involving spin degree of freedom.
	\end{abstract}
	
	\maketitle
	Molecular reactions are driven by the motion and reorganization of valence electrons in an ultrashort time scale~\cite{Fleischer11:PRL107,Woerner10:Nature466,Ischenko17:CR117,Yang20:Science368}. Understanding coherent valence electron dynamics involving spin degrees of freedom in ultrashort time scale is of fundamental importance to the study of spintronics~\cite{Bader10:ARCMP1} and spin dynamics in molecules such as spin crossover and spin transfer~\cite{Auboeck15:NatChem7,Wang17:PRL118}. In the last decades, methods such as attosecond streaking~\cite{Drescher02:Nature419,Schultze10:Science328,Gaumnitz17:OE25}, transient absorption~\cite{Goulielmakis10:Nature466,Yang2016:NatPhoton10}, strong field ionization~\cite{Eckle08:Science322,Popruzhenko14:JPBAMOP47} and high-harmonic generation (HHG)~\cite{Niikura05:PRL94} have been introduced to probe the valence electron wave packets, valence hole migration, and quantum beating between low-lying electronic states in coherent superposition~\cite{Kraus13:PRL111}.
	It is especially intriguing to equip the probe of ultrafast electron dynamics with spin resolution.

	X-ray scattering is a powerful tool to achieve spin resolution. As an indispensable technique in imaging the magnetic structure in condensed matter physics~\cite{Kim12:PRL108}, resonant inelastic x-ray scattering (RIXS) can be combined with phase retrieval (PR) algorithms~\cite{Elser03:JOSAA20,Marchesini07:RSI78,Loh10:PRE82} to attain the structure of the magnetic domain~\cite{Turner11:PRL107,Ashish11:PNAS108}. In the RIXS theory for solids, a solid sample is regarded as multiple ions~\cite{Trammell62:PR126,Hannon69:PR186,Hill96:ACA52}, and the PR algorithm of the scattering pattern will provide values for each ions, which represent the spin magnetic moments at different locations. 
	In the hard x-ray range, non-resonant x-ray scattering is also exploited in different types of materials~\cite{Haverkort07:PRL99,Sokaras12:RSI83}. This technique is often combined with magnetic circular dichroism (MCD) and magnetic linear dichroism (MLD) to study the spin and magnetic properties of materials~\cite{Platzman70:PRB2,Hiraoka15:PRB91,Waterfield16:PRL117}.
	
	In this Letter, we propose that x-ray scattering technique has the capability of spatio-temporally resolving the spin dynamics of valence electrons in the gas phase molecules.
	With aligned gas phase neutral nitric oxide molecules in a coherent wave packet consisting of spin-orbit coupled doublet states, we demonstrate that the circular dichroism (CD) of ultrafast non-resonant hard magnetic x-ray scattering (MXS) can be used to reconstruct the electronic density matrix and to retrieve the transient shape of valence electron orbitals occupied by the unpaired electron. The approach proposed in this work could become an important application of x-ray free electron lasers (XFEL) in probing spin-resolved valence electron dynamics of excited state molecules with ultrafast x-ray diffraction, since XFEL can deliver ultrashort circularly polarized x-ray pulses~\cite{Lutman16}. The ultrafast MXS method is versatile and can be sensitive to the spin dynamics in magnetic as well as non-magnetic molecules.
	
\begin{figure*}[htbp]
		\centering
		\includegraphics[width=\linewidth]{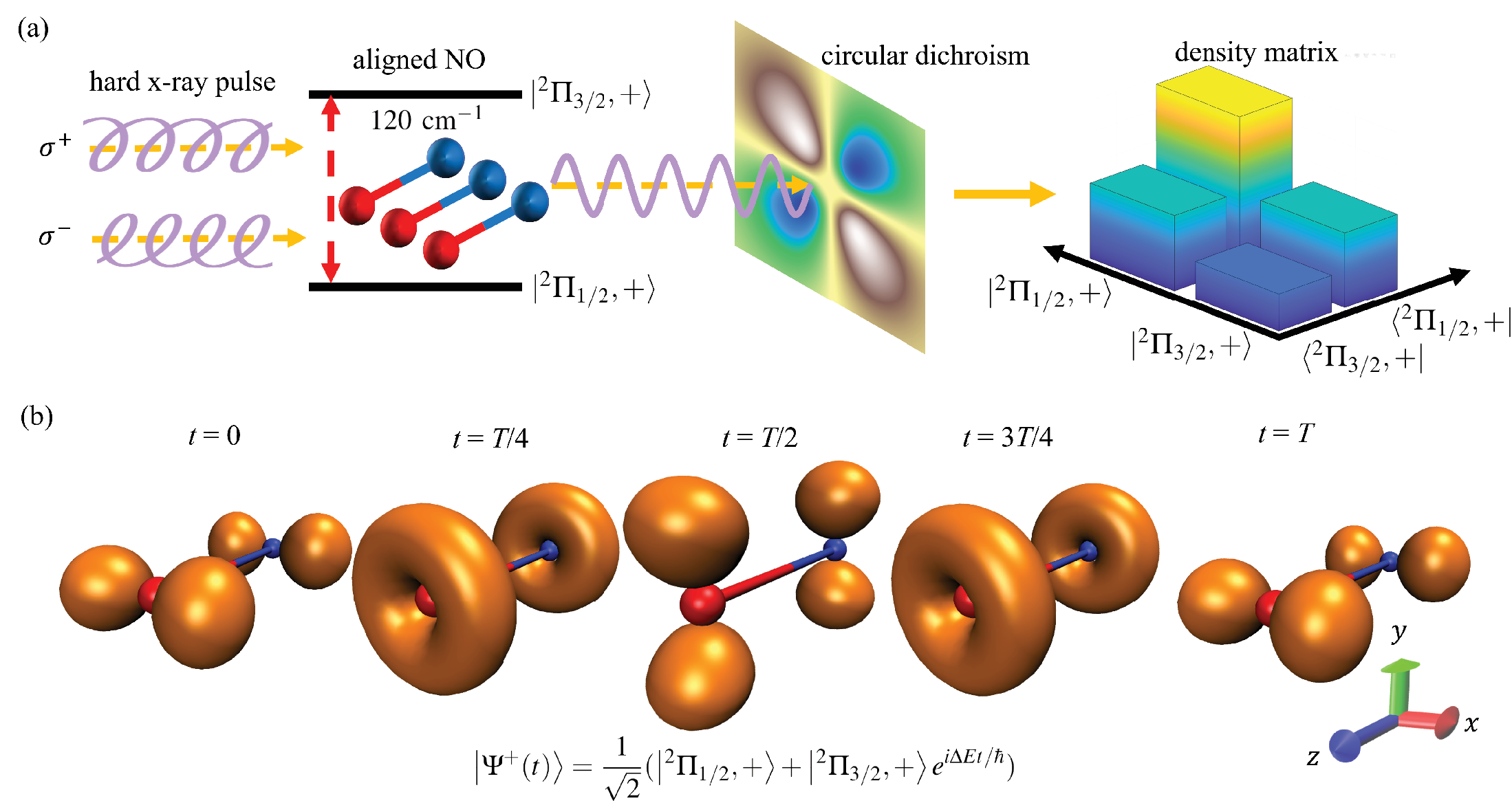}
		\caption{(a) The scheme of reconstructing density matrix in the ultrafast molecular dynamics of $\mathrm{NO}$ from ultrafast non-resonant magnetic x-ray scattering (MXS). First, aligned $\mathrm{NO}$ molecules in the ground state $\ket{ ^2\Pi_{1/2},+}$ are partially excited to $\ket{ ^2\Pi_{3/2},+}$ by stimulated Raman scattering. 
			Then the coherent state is probed by two incident x-ray pulses with right circular ($\sigma^+$) and left circular ($\sigma^-$) polarization to obtain the circular dichroic signal, which cancels the scattering pattern contributed by electric interaction. By comparing the experimental result with ab initio calculation, the off-diagonal elements of the density matrix is finally reconstructed. (b) The time-dependent electron density for the unpaired electron of eq~\ref{eq:Psi+t} for $c_1=1$, which is taken for the purpose of visualization. Red and blue spheres represent the N and O atoms respectively. The spatial orbital of the unpaired electron goes through $\ket{\pi_x}$, $\ket{\pi_x}\pm\ket{\pi_y}$, $\ket{\pi_y}$, $\ket{\pi_x}\mp\ket{\pi_y}$ and $\ket{\pi_x}$ in a complete period $T=278~\mathrm{fs}$ of the quantum beating process. The theoretical derivation for the quantum beating process is in Supplementary Material. }
		\label{fig:scheme}
	\end{figure*}

	As shown in Figure~\ref{fig:scheme}a, a spin-state wave packet of nitric oxide (NO) molecule with an unpaired electron, which is adiabatically aligned with infrared laser and consists of low-lying electronic states $\ket{ ^2\Pi_{1/2}}$ and $\ket{ ^2\Pi_{3/2}}$, can be prepared through impulsive stimulated Raman scattering. The stimulated Raman scattering not only electronically excites the molecules, but also dynamically aligns them as a result of the selection rule.
	The consequent quantum beating proceeds with a period of $T=278~\mathrm{fs}$ that corresponds to energy spacing $\Delta E=120~\mathrm{cm}^{-1}$ of the two states.
	In NO, the transfer from $\ket{ ^2\Pi_{1/2}}$ to $\ket{ ^2\Pi_{3/2}}$ is only weakly allowed, such that $0.1\%$--$0.2\%$ of the molecules can be excited~\cite{lepard70:CJP48,Kraus13:PRL111}.
	The quantum beating is a spin dynamic process since the $\ket{ ^2\Pi_{3/2}}$ has the same spatial electron density as $\ket{ ^2\Pi_{1/2}}$.
	The difference between the two states lies in the spin parts of unpaired electrons, thus it demands an ultrafast spin-resolved probe.
	We demonstrate a method to reconstruct the electronic density matrix from the CD signal of MXS, and the capability of non-resonant MXS in resolving the electron dynamics on both spatial and spin dimensions simultaneously by x-ray scattering patterns of the $\mathrm{NO}$ molecule in coherent spin-state. 
	
	The $\ket{ ^2\Pi_{1/2}}$ and $\ket{ ^2\Pi_{3/2}}$ states are the products of spatial and spin parts, depending on the projection quantum numbers of the electron orbital angular momentum ($\Lambda$) and spin angular momentum ($\Sigma$) along the molecular axis: 
	$\ket{ ^2\Pi_{1/2},+}=(\ket{\pi_+\beta}+\ket{\pi_-\alpha})/\sqrt{2}$ and $\ket{ ^2\Pi_{3/2},+}=(\ket{\pi_+\alpha}+\ket{\pi_-\beta})/\sqrt{2}$. 
	Here the notation $+$ in $\ket{ ^2\Pi_{1/2},+}$ and $\ket{ ^2\Pi_{3/2},+}$ refers to the parity of the electronic levels. The complex-valued spatial states $\ket{\pi^{\pm}}$ represent $\ket{\Lambda=\pm 1}$ and $\ket{\alpha(\beta)}$ represent $\ket{\Sigma=\pm1/2}$.
	They can be further expanded with real-valued spatial orbitals $\ket{\pi_{x}}=(\ket{\pi_{+}}+\ket{\pi_{-}})/\sqrt{2}$ and $\ket{\pi_{y}}=(\ket{\pi_{+}}-\ket{\pi_{-}})/(\sqrt{2}i)$. 
	The unpaired electron density of $\ket{\pi_{x}}$, $\ket{\pi_{\pm}}$ and $\ket{\pi_{y}}$ are shown in Figure~\ref{fig:scheme}b for  $t=0$ $(T)$, $t=T/4$ $(3T/4)$, and $t=T/2$ of the quantum beating, respectively. 
	The stimulated Raman scattering can prepare a coherent superposition state
	\begin{eqnarray} \label{eq:Psi+t}
		\ket{\Psi^+(t)}=N(\ket{ ^2\Pi_{1/2},+}+c_1\ket{ ^2\Pi_{3/2},+}e^{-i\Delta Et/\hbar})\,,
	\end{eqnarray} 
	where $N={1}/({\sqrt{1+|c_1|^2}})$, and the quantum beating can persist for picoseconds~\cite{Kraus13:PRL111}. 
	In non-resonant MXS, the total differential scattering cross section (DSCS) {of an initial electronic state $\ket{i}$} is composed of elastic and inelastic parts (see Sec.~I of Supplementary Material, SM)~\cite{Blume88:PRB37} 
	\begin{eqnarray}
		(\frac{d\sigma}{d\Omega})_{\text{tot}}&=(\frac{e^2}{m_{\mathrm{e}} c^2})^2\bra{i}\sum_{jj'} e^{i\bm{q}\cdot\bm{r}_{j,j'}}\ket{i}\lvert\bm{e}_1\cdot\bm{e}_2^*\rvert^2+(\frac{e^2}{m_{\mathrm{e}} c^2})^2\frac{2\hbar\omega}{m_{\mathrm{e}} c^2}\nonumber\\
		&[\Re\bra{i}\sum_{jj'}e^{i\bm{q}\cdot\bm{r}_{j,j'}}(\frac{c}{\omega}\bm{p}_{j'}\times\frac{\hat{\bm{K}}_1-\hat{\bm{K}_2}}{\hbar})\cdot\bm{P}_1^*(\bm{e}_1\cdot\bm{e}_2^*)\ket{i}\nonumber\\
		&+\Re\bra{i}\sum_{jj'}e^{i\bm{q}\cdot\bm{r}_{j,j'}}i\frac{\bm{\sigma}_{j'}}{2}\cdot\bm{P}_2^*(\bm{e}_1\cdot\bm{e}_2^*)\ket{i}],
		\label{eq:CS}
	\end{eqnarray}
	where $\omega$ is the frequency of incident x-ray, $\bm{q}=\bm{K}_1-\bm{K}_2$ is the momentum transfer of incident x-ray photon,  $\bm{e}_{1}$ and $\bm{e}_{2}$ are the polarization vectors of the incident and outgoing x-ray photon, $\bm{r}_{j,j'}=\bm{r}_{j}-\bm{r}_{j'}$ is the coordinate difference between the $j$th and $j'$th electrons, $\bm{p}_{j'}$ and $\bm{\sigma}_{j'}$ are the momentum and the spin operators of the $j'$th electron.
	The two polarization factors $\bm{P}_1=\bm{e}_1\times\bm{e}_2^*$ and $\bm{P}_2=\bm{e}_2^*\times\bm{e}_1-(\hat{\bm{K}}_2\times\bm{e}_2^*)\times(\hat{\bm{K}}_1\times\bm{e}_1)-(\bm{e}_2^*\cdot\hat{\bm{K}}_1)(\hat{\bm{K}}_1\times\bm{e}_1)+(\bm{e}_1\cdot
	\hat{\bm{K}}_2)(\hat{\bm{K}}_2\times\bm{e}_2^*)$ only rely on the x-ray parameters. 
	
	The DSCS in eq~\ref{eq:CS} consists of the contribution from Thomson charge scattering, where $\sum_{jj'} e^{i\bm{q}\cdot\bm{r}_{j,j'}}$ indicates the two-electron correlation density function~\cite{Bartell64:JACS86}. 
	The second and the third terms represent the orbital and spin-dependent contributions~\cite{Blume85:JAP57}.
	For hard x-ray of $0.2$~\AA~wavelength, the spin-dependent cross section is $\frac{\hbar\omega}{N_{\mathrm{e}}m_{\mathrm{e}} c^2}\approx 9\times10^{-3}$ smaller than the charge scattering, where $N_{\mathrm{e}}$ is the number of electrons in one molecule and $\hbar\omega$ is the energy of incident x-ray photon. 
	The charge scattering contribution in DSCS can be eliminated by probing the circular dichroism signal~\cite{BookXrayScatter}.

\begin{figure}[htbp]
		\centering
		\includegraphics[width=\linewidth]{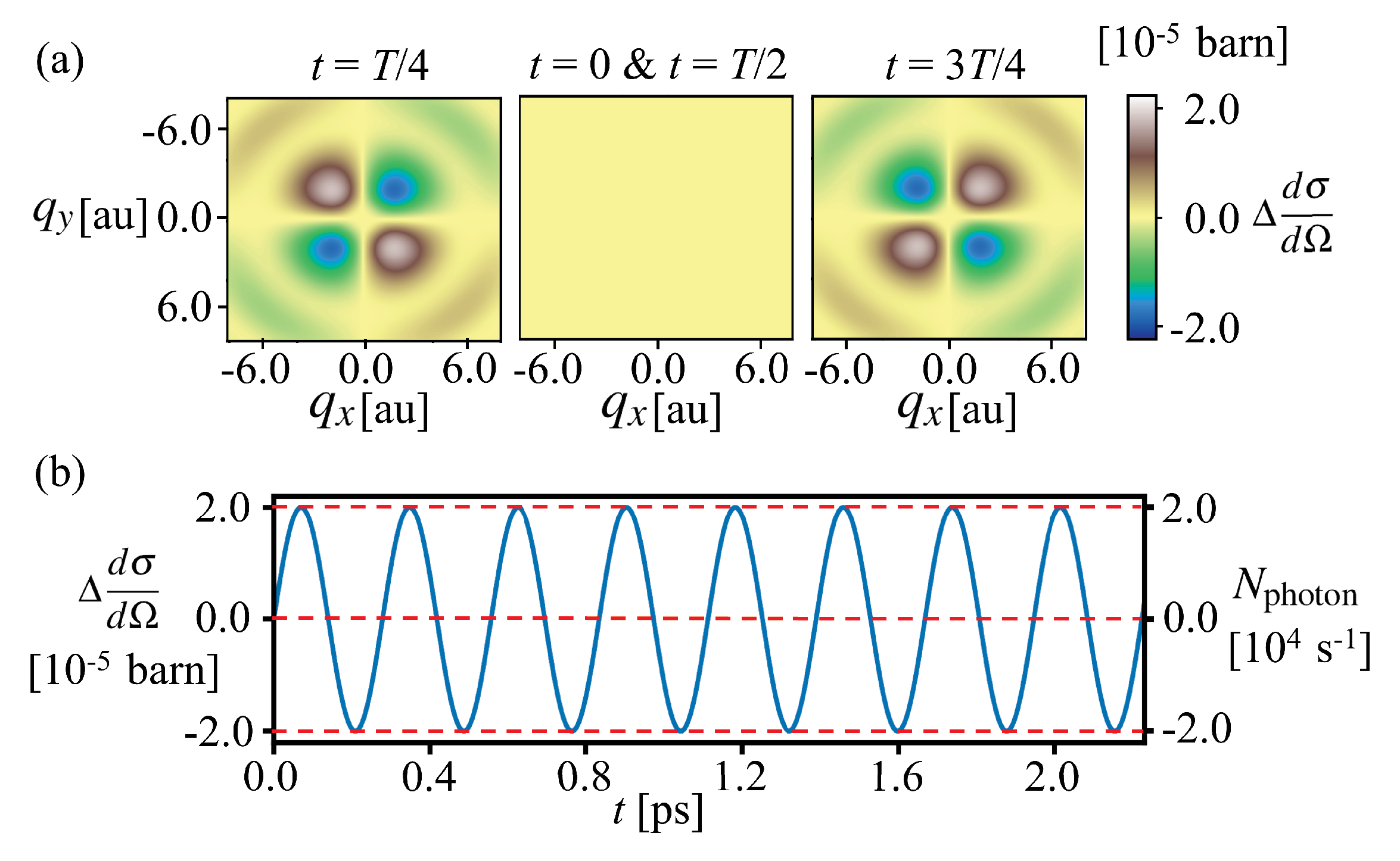}
		\caption{(a) Simulation of ultrafast MXS circular dichroism pattern at time $t=T/4$, $t=0~ \&~t=T/2$ and $t=3T/4$ for perfectly aligned NO molecules. 
			The superposition state is chosen as $\ket{\Psi^+(t)}=\frac{1}{\sqrt{2}}(\ket{ ^2\Pi_{1/2},+}+\ket{ ^2\Pi_{3/2},+}e^{-i\Delta Et/\hbar})$. The incident X-ray is along the $\mathrm{NO}$ axis and has a wavelength of $0.2$~\AA. (b) The time-dependent differential scattering cross section (DSCS) $\Delta\frac{d\sigma}{d\Omega}$ for the perfectly aligned molecules and the corresponding signal strength in photon number ($N_{\mathrm{photon}}$) per unit solid angle at position where DSCS reaches its maximum. $N_{\mathrm{photon}}$ is estimated with XFEL parameters in Supplementary Material.}
		\label{fig:beating}
	\end{figure}

	The ultrafast non-resonant MXS of spin-state quantum beating is investigated using wave function from the ab initio calculation at restricted active space self-consistent field (RASSCF) level~\cite{Malmqvist90:JPC94}.
	The electronic wave function $\Psi_k^{(S,M_S)}$ is represented in the basis of configuration state functions (CSFs) $\ket{\Phi_k^{(S,M_S)}}$:
	\begin{align}\label{eq:CSF}
		\ket{\Psi(t)}=\sum_{k}a_k(t)\ket{\Phi_k^{(S,M_S)}},
	\end{align}
	with the total spin $S$ and its projection $M_S$ on the molecular axis. 
	The corresponding magnetic x-ray scattering patterns from the aligned NO molecules in the coherent superposition of spin-states are simulated using eq~\ref{eq:CS} and wave function in the CSF basis (see Sec.~II of SM for the ab initio calculation of MXS from the coherent spin-state wave packet).
	In Figure~\ref{fig:beating}a, the dichroic signals are shown for the MXS from the NO molecules in the spin-state $\ket{\Psi^+(t)}=(\ket{ ^2\Pi_{1/2},+}+\ket{ ^2\Pi_{3/2},+}e^{-i\Delta Et/\hbar})/\sqrt{2}$ within a period of quantum beating. 
	In this case, the orbital contribution of the magnetic scattering is proved to be zero with ab initio calculation. Thus the dichroic patterns only depend on the spin-dependent contribution of the magnetic scattering.
	The CD signal is dominated by the transitions between CSFs $\ket{\Phi_k^{(S,M_S)}}$ of same spin projection $M_S$ in the x-ray scattering.
	The transitions between CSFs of different spin projections through $\sigma_{\pm}$ don't contribute to the circular dichroism 
	because the molecule has cylindrical symmetry around the NO axis (see Sec.~IV of SM for details of vanishing MXS signal contributed from $\sigma_{\pm}$).
	In real space, the alignment of NO molecular axis with an impulsive laser is non-perfect.  
	The alignment degree can reach $\langle \cos^2\eta \rangle\sim 0.4$--$0.6$~\cite{Kraus13:PRL111} for the angle $\eta$ between NO axis and the $Z$ axis of lab frame, which is perpendicular to the laser linear polarization direction.
	And the initial excitation acts selectively on the electron in the $\pi_x$ orbital, which aligns parallel with the laser polarization.

	In Figure~\ref{fig:beating}a, the petaloid-shaped patterns at time $t=T/4$ and $t=3T/4$ possess nodes along the $q_x$ and $q_y$ axes, it reflects spatial symmetry of the transition matrix element $\bra{\pi_x}S_{\mathrm{R}}\ket{\pi_y}$, where $S_{\mathrm{R}}$ is the real part of the scattering operator $S=\sum_{jj'}e^{i\bm{q}\cdot\bm{r}_{j,j'}}i\frac{\bm{\sigma}_{j'}}{2}\cdot\Delta[\bm{P}_2^*(\bm{e}_1\cdot\bm{e}_2^*)]$, and $\Delta[\bm{P}_2^*(\bm{e}_1\cdot\bm{e}_2^*)]$ is the polarization factor in the circular dichroic signal, as shown in Sec.~I and III of SM.
	In eq~\ref{eq:dcs(t)} and Figure~\ref{fig:beating}a, the selective $\pi_x$ orbital excitation for the dynamically aligned molecule avoids the averaging over $\pi_x$ and $\pi_y$ orbitals, otherwise MXS CD signal could vanish.
	In this case, the operator $S$ in eq~\ref{eq:CS} has non-zero amplitude only between CSFs with the same $\pi_x$ or $\pi_y$ orbital occupation.
	In fact, it can be proved that the time-dependent DSCS between the two circularly polarized x-ray of eq~\ref{eq:Psi+t} is (see Sec.~V in SM)
	\begin{align}
		\Delta\frac{d\sigma}{d\Omega}(\bm{q},t)=\left\{\begin{aligned}
			&0 &t<0\\
			&\frac{2|c_1|\bra{\pi_y\alpha}S_{\mathrm{R}}(\bm{q})\ket{\pi_x\alpha}}{1+|c_1|^2}\sin(\omega_0 t-\varphi)&t\ge0
		\end{aligned}\right.,\label{eq:dcs(t)}
	\end{align}
	where $c_1=|c_1|e^{i\varphi}$ is the complex amplitude in eq~\ref{eq:Psi+t} and $\omega_0=2\pi/T$ is the frequency of quantum beating.
	eq~\ref{eq:dcs(t)} implies the oscillatory cross section presented in Figure~\ref{fig:beating}b.
	From the practical perspective, we also estimated photon number counts of MXS circular dichroism.
	Typically $0.1\%$--$0.2\%$ of the NO molecules are electronically excited in the stimulated Raman scattering~\cite{Kraus13:PRL111}. Thus we assume $|c_1|=\sqrt{0.002}$, and the maximum scattering photon number per second in Figure~\ref{fig:beating}b is estimated to be $N_{\mathrm{photon},0.2\%}=1.8\times10^3~\mathrm{s^{-1}}$, as shown in Sec.~VI of SM.
	For XFEL in the self-amplified spontaneous emission (SASE)~\cite{Emma10:NatPhoton4} or seeded~\cite{Lambert08:NatPhys4} regime, the light intensity fluctuation can introduce error between the left and right circularly polarized x-ray pulses. To balance the light intensity for different pulses, we suggest repeated experiment as a necessity for MXS circular dichroism.
	For an experiment with unknown $c_1$ in $\mathrm{NO}$ superposition state, the ab initio calculated $\bra{\pi_y\alpha}S_{\mathrm{R}}(\bm{q})\ket{\pi_x\alpha}$ can be used to retrieve $|c_1|$ and $\varphi$ from the observed time-dependent MXS CD signal.
	During long time evolution, quantum decoherence 
	leads to a decreasing $|\rho_{12}|$, where $\ket{1}$ and $\ket{2}$ represent $\ket{\Pi_{1/2},+}$ and $\ket{\Pi_{3/2},+}$. 
	In principle, one can reconstruct the off-diagonal term in the density matrix of $\ket{\Pi_{1/2(3/2)},+}$ with MXS in the case with quantum decoherence.
	
	\begin{figure}[htbp]
		\centering
		\includegraphics[width=\linewidth]{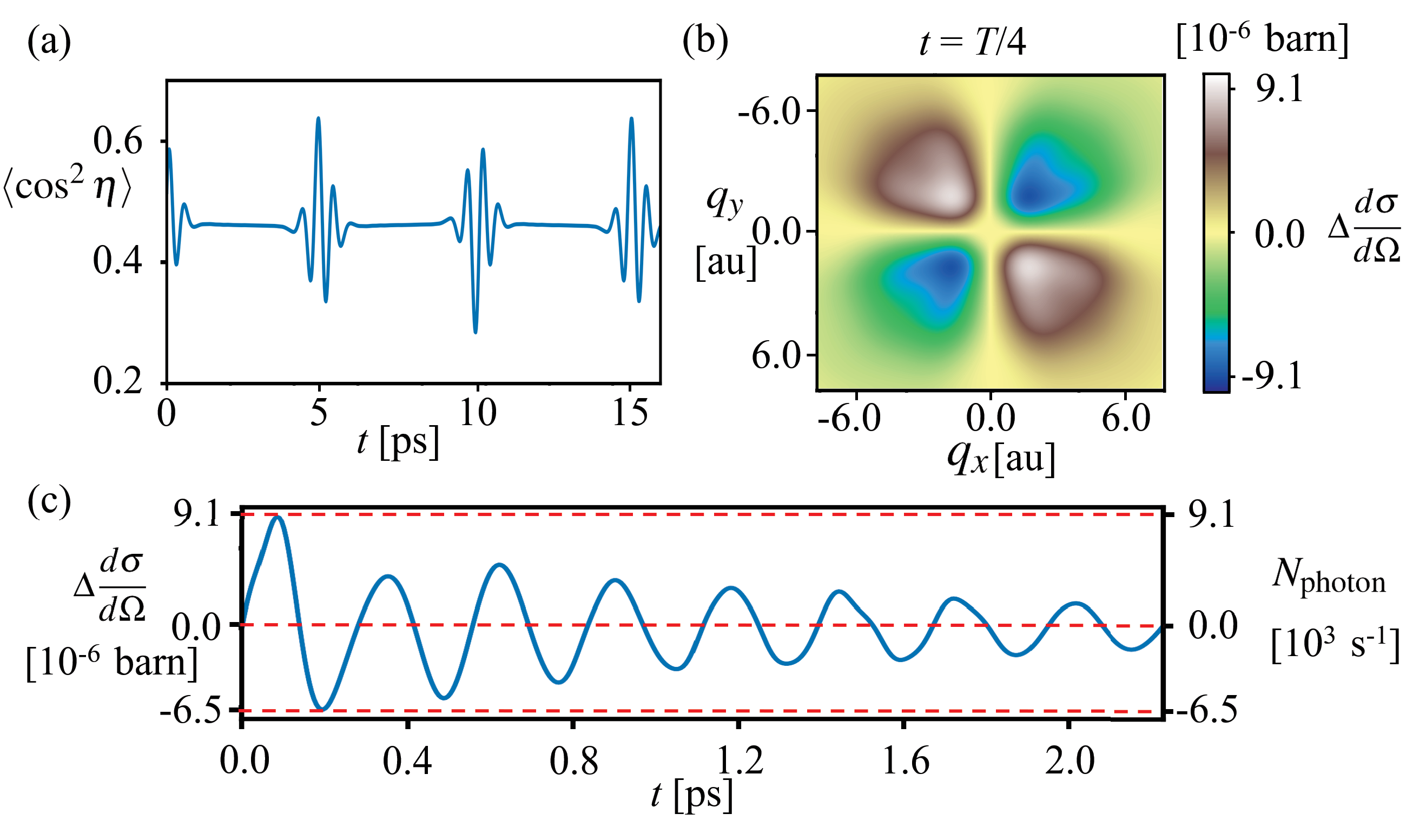}
		\caption{(a) The degree of the molecular alignment by solving time-dependent Schr\"{o}dinger equation. The NO molecules are at the temperature $20$~K. The 800~nm alignment pulse has a full width half maximum (FWHM) of $60$~fs and the peak intensity is $6\times 10^{13}$~W/cm$^2$.
			(b) Simulated ultrafast MXS circular dichroism pattern at time $t=T/4$ for the model with non-perfect alignment and decoherence. The incident x-ray is along the $\mathrm{NO}$ axis and has a wavelength of $0.2$~\AA. 
			(c) The time-dependent DSCS $\Delta\frac{d\sigma}{d\Omega}$ and the corresponding signal strength in photon number ($N_{\mathrm{photon}}$) per unit solid angle at position where DSCS reaches its maximum. 
			The damped amplitude of the oscillatory MXS CD signal exhibits the influence of quantum decoherence.
		}
		\label{fig:non-perfect}
	\end{figure}

	Non-perfect alignment of the NO rotational wave packet is a non-negligible effect, and the simulated MXS CD signal involving non-perfect alignment as well as the quantum decoherence is shown in Figure~\ref{fig:non-perfect}.
	By solving time-dependent Schr\"{o}dinger equation, the degree of the molecular alignment $\langle\cos^2\eta\rangle(t)$ in Figure~\ref{fig:non-perfect}a is calculated for an 800~nm alignment pulse of $60$~fs FWHM and the peak intensity is $6\times 10^{13}$~W/cm$^2$. The time interval between revivals of coupled rotational and electronic wave packet is about $5$~ps.
	Figure~\ref{fig:non-perfect}b shows the MXS CD signal at time $t=T/4$, and for other time $t$, the circular dichroic signal exhibits similar oscillatory pattern but decaying intensity. 
	Non-perfect alignment of molecules leads to diffusion of the rotationally averaged spin density in the laboratory frame.
	Thus the MXS CD pattern exhibits larger petaloid area than that in Figure~\ref{fig:beating}a when perfect alignment is considered. Compared with the case of perfect alignment, Figure~\ref{fig:non-perfect}b has lower intensity since the averaged expectation of the scattering operator $S$ is affected by $\cos\eta$ and quantum decoherence.

	In Figure~\ref{fig:non-perfect}c, we present the MXS CD signal at position where DSCS reaches its maximum in the reciprocal space.
	The MXS CD signal has an approximate form of a decaying sine function for the time period between the rotational revivals.
	According to Sec.~V in SM, the decaying sine function of $\Delta\frac{d\sigma}{d\Omega}$  is proportional to off-diagonal density matrix elements. 
	By comparing the value of this decaying sine function with the simulation under the same degree of alignment, it is possible to estimate the off-diagonal density matrix elements in the presence of non-perfect alignment.

\begin{figure}[htbp]
		\centering
		\includegraphics[width=\linewidth]{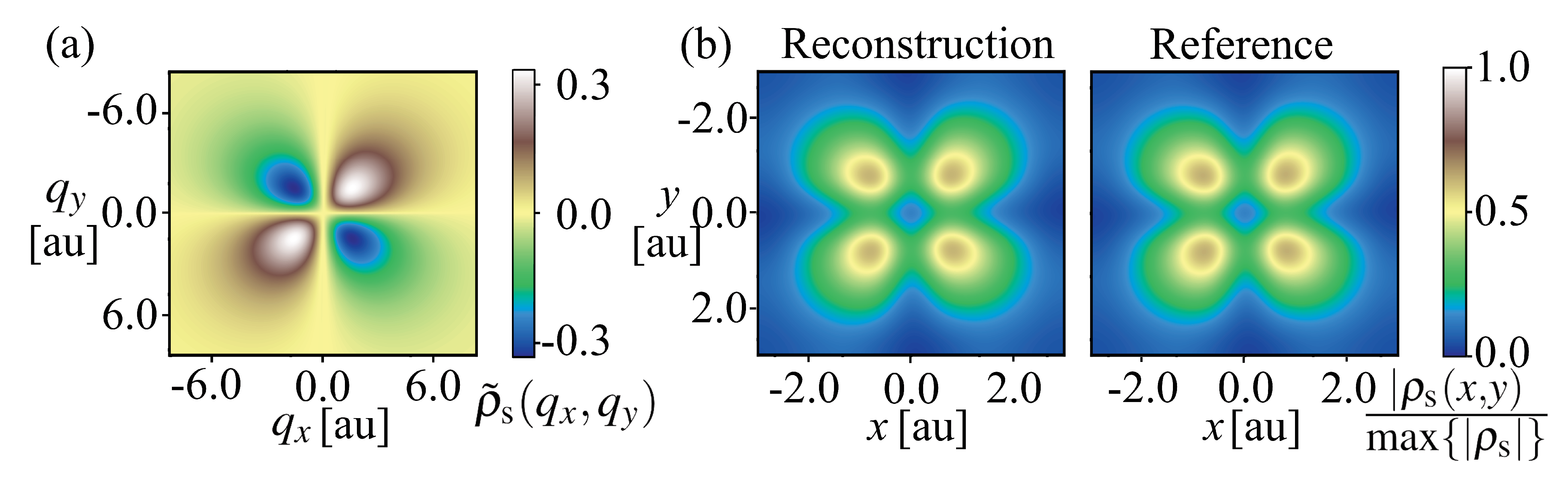}
		\caption{Spin density of NO molecule retrieved from MXS dichroic signal in Figure~\ref{fig:beating}a at $t=T/4$.
			(a) 2-dimensional spin density $\tilde{\rho}_{\mathrm{s}}(q_x,q_y)$ in the momentum space.
			(b) Normalized 2-dimensional spin density $|\rho_{\mathrm{s}}(x,y)|/\max\{|\rho_{\mathrm{s}}|\}$ in the real space. The spin density is distributed in the range of $[-2.0,2.0]$~au in both $x$ and $y$ direction. %
			The left one of $|\rho_{\mathrm{s}}(x,y)|/\max\{|\rho_{\mathrm{s}}|\}$ is reconstructed from MXS dichroic signal, the right one is directly calculated
			for reference.
		}
		\label{fig:spindensity}
	\end{figure}

	Apart from reconstructing the evolving density matrix in ultrafast molecular spin dynamics, MXS also shows the capability in resolving spatial distribution of spin density.
	In the quantum beating process of Figure~\ref{fig:scheme}b, the spin density is defined as the difference of the densities of electrons with $\alpha$ and $\beta$ spin
	\begin{align}\label{eq:spindensity}
		\rho_{\mathrm{s}}(\bm{r})
		=&\rho_{\alpha}(\bm{r})-\rho_{\beta}(\bm{r}) \notag\\
		=&\bra{\Psi^+(t)}\sum_{j}\delta(\bm{r}-\bm{r}_j)\bm{\sigma}_j/2\ket{\Psi^+(t)} \notag\\
		=&\bra{\pi_y}\sum_{j}\delta(\bm{r}-\bm{r}_j)/2\ket{\pi_x}\sin(\omega_0 t),
	\end{align}
	where $j$ is the index of electrons. 
	According to eq~\ref{eq:spindensity}, the spin density only changes its amplitude in the quantum beating process, but stays the same in shape.
	In this case, only the $z$-component of the spin density is non-zero, thus we will denote $\rho_{\mathrm{s}}(\bm{r})$ as its $z$-component without ambiguity in the following statements.
	As discussed in the Sec.~II of SM, the MXS dichroic signal can be approximately viewed as the product of electron density of the inner shell electrons and the total spin density in the momentum space. 
	We retrieve the 2-dimensional spin density $\tilde{\rho}_{\mathrm{s}}(q_x,q_y)$ in the reciprocal space, as shown in Figure~\ref{fig:spindensity}a.
	With Fourier transformation of $q_x$ and $q_y$, the normalized 2-dimensional spin density $|\rho_{\mathrm{s}}(x,y)|/\max\{|\rho_{\mathrm{s}}|\}$ in the real space is calculated in Figure~\ref{fig:spindensity}b. 
	In Figure~\ref{fig:spindensity}b, the 2-dimensional spin density is consistent with eq~\ref{eq:spindensity}, which shows the spatial overlap between the unpaired electrons of $\ket{\pi_x}$ and $\ket{\pi_y}$.
	By comparing the reconstructed spin density with the reference one, it shows the capability of MXS dichroism in probing the spin density with high accuracy.

	To conclude, we propose an approach to probe the ultrafast spin dynamics in molecules and apply it to the aligned $\mathrm{NO}$ molecules in the simulation. 
	The circular dichroism of MXS can eliminate the charge scattering signal between the incoming x-ray and molecules, and probes electrons via the magnetic interaction.
	We have proposed that by fitting the experimental MXS result with ab initio calculation, MXS circular dichroism can reconstruct the electronic density matrix on ultrafast time scale. 
	We have also shown that MXS is capable of probing the spin density by resolving the spatial distribution.
	The technique is promising in probing molecular dynamics since it provides information about spin-polarized electronic with high temporal and spatial resolution. Given the recent progress of XFEL approaching the attosecond regime~\cite{Emma10:NatPhoton4,Huang17:PRL119} and x-ray pulses with circular polarization~\cite{Lutman16:NatPhoton10}.
	The experimental implementation of the ultrafast MXS probe for electron spin dynamics is within reach.

	\section{SUPPORTING INFORMATION}
	theoretical derivation of differential cross section, calculation of NO molecular dynamics and magnetic x-ray scattering signal, photon number rate estimation
	\section*{ACKNOWLEDGMENTS}
	We gratefully acknowledge Haitan Xu, R. J. Dwayne Miller, Oriol Vendrell and Linfeng Zhang for helpful discussions.
	This work has been supported by National Natural Science Foundation of China (Grants No.~12234002, 12174009, 92250303), and Beijing Natural Science Foundation (Grant No.~Z220008).

	\providecommand{\latin}[1]{#1}
	\makeatletter
	\providecommand{\doi}
	{\begingroup\let\do\@makeother\dospecials
		\catcode`\{=1 \catcode`\}=2 \doi@aux}
	\providecommand{\doi@aux}[1]{\endgroup\texttt{#1}}
	\makeatother
	\providecommand*\mcitethebibliography{\thebibliography}
	\csname @ifundefined\endcsname{endmcitethebibliography}
	{\let\endmcitethebibliography\endthebibliography}{}


\begin{mcitethebibliography}{42}
		\providecommand*\natexlab[1]{#1}
		\providecommand*\mciteSetBstSublistMode[1]{}
		\providecommand*\mciteSetBstMaxWidthForm[2]{}
		\providecommand*\mciteBstWouldAddEndPuncttrue
		{\def\EndOfBibitem{\unskip.}}
		\providecommand*\mciteBstWouldAddEndPunctfalse
		{\let\EndOfBibitem\relax}
		\providecommand*\mciteSetBstMidEndSepPunct[3]{}
		\providecommand*\mciteSetBstSublistLabelBeginEnd[3]{}
		\providecommand*\EndOfBibitem{}
		\mciteSetBstSublistMode{f}
		\mciteSetBstMaxWidthForm{subitem}{(\alph{mcitesubitemcount})}
		\mciteSetBstSublistLabelBeginEnd
		{\mcitemaxwidthsubitemform\space}
		{\relax}
		{\relax}
		
		\bibitem[Fleischer \latin{et~al.}(2011)Fleischer, W\"orner, Arissian, Liu,
		Meckel, Rippert, D\"orner, Villeneuve, Corkum, and
		Staudte]{Fleischer11:PRL107}
		Fleischer,~A.; W\"orner,~H.~J.; Arissian,~L.; Liu,~L.~R.; Meckel,~M.;
		Rippert,~A.; D\"orner,~R.; Villeneuve,~D.~M.; Corkum,~P.~B.; Staudte,~A.
		Probing Angular Correlations in Sequential Double Ionization. \emph{Phys.
			Rev. Lett.} \textbf{2011}, \emph{107}, 113003\relax
		\mciteBstWouldAddEndPuncttrue
		\mciteSetBstMidEndSepPunct{\mcitedefaultmidpunct}
		{\mcitedefaultendpunct}{\mcitedefaultseppunct}\relax
		\EndOfBibitem
		\bibitem[Woerner \latin{et~al.}(2010)Woerner, Bertrand, Kartashov, Corkum, and
		Villeneuve]{Woerner10:Nature466}
		Woerner,~H.~J.; Bertrand,~J.~B.; Kartashov,~D.~V.; Corkum,~P.~B.;
		Villeneuve,~D.~M. Following a Chemical Reaction Using High-harmonic
		Interferometry. \emph{Nature} \textbf{2010}, \emph{466}, 604--607\relax
		\mciteBstWouldAddEndPuncttrue
		\mciteSetBstMidEndSepPunct{\mcitedefaultmidpunct}
		{\mcitedefaultendpunct}{\mcitedefaultseppunct}\relax
		\EndOfBibitem
		\bibitem[Ischenko \latin{et~al.}(2017)Ischenko, Weber, and
		Miller]{Ischenko17:CR117}
		Ischenko,~A.~A.; Weber,~P.~M.; Miller,~R. J.~D. Capturing Chemistry in Action
		with Electrons: Realization of Atomically Resolved Reaction Dynamics.
		\emph{Chem. Rev.} \textbf{2017}, \emph{117}, 11066--11124\relax
		\mciteBstWouldAddEndPuncttrue
		\mciteSetBstMidEndSepPunct{\mcitedefaultmidpunct}
		{\mcitedefaultendpunct}{\mcitedefaultseppunct}\relax
		\EndOfBibitem
		\bibitem[Yang \latin{et~al.}(2020)Yang, Zhu, Nunes, Yu, Parrish, Wolf,
		Centurion, Gühr, Li, Liu, Moore, Niebuhr, Par, Shen, Weathersby, Weinacht,
		Martinez, and Wang]{Yang20:Science368}
		Yang,~J.; Zhu,~X.; Nunes,~J. P.~F.; Yu,~J.~K.; Parrish,~R.~M.; Wolf,~T. J.~A.;
		Centurion,~M.; Gühr,~M.; Li,~R.; Liu,~Y. \latin{et~al.}  Simultaneous
		Observation of Nuclear and Electronic Dynamics by Ultrafast Electron
		Diffraction. \emph{Science} \textbf{2020}, \emph{368}, 885--889\relax
		\mciteBstWouldAddEndPuncttrue
		\mciteSetBstMidEndSepPunct{\mcitedefaultmidpunct}
		{\mcitedefaultendpunct}{\mcitedefaultseppunct}\relax
		\EndOfBibitem
		\bibitem[Bader and Parkin(2010)Bader, and Parkin]{Bader10:ARCMP1}
		Bader,~S.; Parkin,~S. Spintronics. \emph{Annu. Rev. Condens. Matter Phys.}
		\textbf{2010}, \emph{1}, 71--88\relax
		\mciteBstWouldAddEndPuncttrue
		\mciteSetBstMidEndSepPunct{\mcitedefaultmidpunct}
		{\mcitedefaultendpunct}{\mcitedefaultseppunct}\relax
		\EndOfBibitem
		\bibitem[Auboeck and Chergui(2015)Auboeck, and Chergui]{Auboeck15:NatChem7}
		Auboeck,~G.; Chergui,~M. Sub-$50$-fs Photoinduced Spin Crossover in
		$[\mathrm{Fe(bpy)}_{3}]^{2+}$. \emph{Nat. Chem.} \textbf{2015}, \emph{7},
		629--633\relax
		\mciteBstWouldAddEndPuncttrue
		\mciteSetBstMidEndSepPunct{\mcitedefaultmidpunct}
		{\mcitedefaultendpunct}{\mcitedefaultseppunct}\relax
		\EndOfBibitem
		\bibitem[Wang \latin{et~al.}(2017)Wang, Bokarev, Aziz, and
		K\"uhn]{Wang17:PRL118}
		Wang,~H.; Bokarev,~S.~I.; Aziz,~S.~G.; K\"uhn,~O. Ultrafast Spin-State Dynamics
		in Transition-Metal Complexes Triggered by Soft-x-ray Light. \emph{Phys. Rev.
			Lett.} \textbf{2017}, \emph{118}, 023001\relax
		\mciteBstWouldAddEndPuncttrue
		\mciteSetBstMidEndSepPunct{\mcitedefaultmidpunct}
		{\mcitedefaultendpunct}{\mcitedefaultseppunct}\relax
		\EndOfBibitem
		\bibitem[Drescher \latin{et~al.}(2002)Drescher, Hentschel, Kienberger,
		Uiberacker, Yakovlev, Scrinzi, Westerwalbesloh, Kleineberg, Heinzmann, and
		Krausz]{Drescher02:Nature419}
		Drescher,~M.; Hentschel,~M.; Kienberger,~R.; Uiberacker,~M.; Yakovlev,~V.;
		Scrinzi,~A.; Westerwalbesloh,~T.; Kleineberg,~U.; Heinzmann,~U.; Krausz,~F.
		Time-resolved Atomic Inner-shell Spectroscopy. \emph{Nature} \textbf{2002},
		\emph{419}, 803--807\relax
		\mciteBstWouldAddEndPuncttrue
		\mciteSetBstMidEndSepPunct{\mcitedefaultmidpunct}
		{\mcitedefaultendpunct}{\mcitedefaultseppunct}\relax
		\EndOfBibitem
		\bibitem[Schultze \latin{et~al.}(2010)Schultze, Fiess, Karpowicz, Gagnon,
		Korbman, Hofstetter, Neppl, Cavalieri, Komninos, Mercouris, Nicolaides,
		Pazourek, Nagele, Feist, Burgdörfer, Azzeer, Ernstorfer, Kienberger,
		Kleineberg, Goulielmakis, Krausz, and Yakovlev]{Schultze10:Science328}
		Schultze,~M.; Fiess,~M.; Karpowicz,~N.; Gagnon,~J.; Korbman,~M.;
		Hofstetter,~M.; Neppl,~S.; Cavalieri,~A.~L.; Komninos,~Y.; Mercouris,~T.
		\latin{et~al.}  Delay in Photoemission. \emph{Science} \textbf{2010},
		\emph{328}, 1658--1662\relax
		\mciteBstWouldAddEndPuncttrue
		\mciteSetBstMidEndSepPunct{\mcitedefaultmidpunct}
		{\mcitedefaultendpunct}{\mcitedefaultseppunct}\relax
		\EndOfBibitem
		\bibitem[Gaumnitz \latin{et~al.}(2017)Gaumnitz, Jain, Pertot, Huppert, Jordan,
		Ardana-Lamas, and Worner]{Gaumnitz17:OE25}
		Gaumnitz,~T.; Jain,~A.; Pertot,~Y.; Huppert,~M.; Jordan,~I.; Ardana-Lamas,~F.;
		Worner,~H.~J. Streaking of $43$-attosecond Soft-x-ray Pulses Generated by a
		Passively CEP-stable Mid-infrared Driver. \emph{Opt. Express} \textbf{2017},
		\emph{25}, 27506--27518\relax
		\mciteBstWouldAddEndPuncttrue
		\mciteSetBstMidEndSepPunct{\mcitedefaultmidpunct}
		{\mcitedefaultendpunct}{\mcitedefaultseppunct}\relax
		\EndOfBibitem
		\bibitem[Goulielmakis \latin{et~al.}(2010)Goulielmakis, Loh, Wirth, Santra,
		Rohringer, Yakovlev, Zherebtsov, Pfeifer, Azzeer, Kling, Leone, and
		Krausz]{Goulielmakis10:Nature466}
		Goulielmakis,~E.; Loh,~Z.-H.; Wirth,~A.; Santra,~R.; Rohringer,~N.;
		Yakovlev,~V.~S.; Zherebtsov,~S.; Pfeifer,~T.; Azzeer,~A.~M.; Kling,~M.~F.
		\latin{et~al.}  Real-time Observation of Valence Electron Motion.
		\emph{Nature} \textbf{2010}, \emph{466}, 739--743\relax
		\mciteBstWouldAddEndPuncttrue
		\mciteSetBstMidEndSepPunct{\mcitedefaultmidpunct}
		{\mcitedefaultendpunct}{\mcitedefaultseppunct}\relax
		\EndOfBibitem
		\bibitem[Yang \latin{et~al.}(2016)Yang, Ostrowski, France, Zhu, van~de
		Lagemaat, Luther, and Beard]{Yang2016:NatPhoton10}
		Yang,~Y.; Ostrowski,~D.~P.; France,~R.~M.; Zhu,~K.; van~de Lagemaat,~J.;
		Luther,~J.~M.; Beard,~M.~C. Observation of a Hot-phonon Bottleneck in
		Lead-iodide Perovskites. \emph{Nat. Photon.} \textbf{2016}, \emph{10},
		53--59\relax
		\mciteBstWouldAddEndPuncttrue
		\mciteSetBstMidEndSepPunct{\mcitedefaultmidpunct}
		{\mcitedefaultendpunct}{\mcitedefaultseppunct}\relax
		\EndOfBibitem
		\bibitem[Eckle \latin{et~al.}(2008)Eckle, Pfeiffer, Cirelli, Staudte, Dörner,
		Muller, Büttiker, and Keller]{Eckle08:Science322}
		Eckle,~P.; Pfeiffer,~A.~N.; Cirelli,~C.; Staudte,~A.; Dörner,~R.;
		Muller,~H.~G.; Büttiker,~M.; Keller,~U. Attosecond Ionization and Tunneling
		Delay Time Measurements in Helium. \emph{Science} \textbf{2008}, \emph{322},
		1525--1529\relax
		\mciteBstWouldAddEndPuncttrue
		\mciteSetBstMidEndSepPunct{\mcitedefaultmidpunct}
		{\mcitedefaultendpunct}{\mcitedefaultseppunct}\relax
		\EndOfBibitem
		\bibitem[Popruzhenko(2014)]{Popruzhenko14:JPBAMOP47}
		Popruzhenko,~S.~V. Keldysh Theory of Strong Field Ionization: History,
		Applications, Difficulties and Perspectives. \emph{J. Phys. B: At. Mol. Opt.
			Phys.} \textbf{2014}, \emph{47}, 204001\relax
		\mciteBstWouldAddEndPuncttrue
		\mciteSetBstMidEndSepPunct{\mcitedefaultmidpunct}
		{\mcitedefaultendpunct}{\mcitedefaultseppunct}\relax
		\EndOfBibitem
		\bibitem[Niikura \latin{et~al.}(2005)Niikura, Villeneuve, and
		Corkum]{Niikura05:PRL94}
		Niikura,~H.; Villeneuve,~D.~M.; Corkum,~P.~B. Mapping Attosecond Electron Wave
		Packet Motion. \emph{Phys. Rev. Lett.} \textbf{2005}, \emph{94}, 083003\relax
		\mciteBstWouldAddEndPuncttrue
		\mciteSetBstMidEndSepPunct{\mcitedefaultmidpunct}
		{\mcitedefaultendpunct}{\mcitedefaultseppunct}\relax
		\EndOfBibitem
		\bibitem[Kraus \latin{et~al.}(2013)Kraus, Zhang, Gijsbertsen, Lucchese,
		Rohringer, and W\"orner]{Kraus13:PRL111}
		Kraus,~P.~M.; Zhang,~S.~B.; Gijsbertsen,~A.; Lucchese,~R.~R.; Rohringer,~N.;
		W\"orner,~H.~J. High-Harmonic Probing of Electronic Coherence in Dynamically
		Aligned Molecules. \emph{Phys. Rev. Lett.} \textbf{2013}, \emph{111},
		243005\relax
		\mciteBstWouldAddEndPuncttrue
		\mciteSetBstMidEndSepPunct{\mcitedefaultmidpunct}
		{\mcitedefaultendpunct}{\mcitedefaultseppunct}\relax
		\EndOfBibitem
		\bibitem[Kim \latin{et~al.}(2012)Kim, Casa, Upton, Gog, Kim, Mitchell, van
		Veenendaal, Daghofer, van~den Brink, Khaliullin, and Kim]{Kim12:PRL108}
		Kim,~J.; Casa,~D.; Upton,~M.~H.; Gog,~T.; Kim,~Y.-J.; Mitchell,~J.~F.; van
		Veenendaal,~M.; Daghofer,~M.; van~den Brink,~J.; Khaliullin,~G.
		\latin{et~al.}  Magnetic Excitation Spectra of
		${\mathrm{Sr}}_{2}{\mathrm{IrO}}_{4}$ Probed by Resonant Inelastic X-Ray
		Scattering: Establishing Links to Cuprate Superconductors. \emph{Phys. Rev.
			Lett.} \textbf{2012}, \emph{108}, 177003\relax
		\mciteBstWouldAddEndPuncttrue
		\mciteSetBstMidEndSepPunct{\mcitedefaultmidpunct}
		{\mcitedefaultendpunct}{\mcitedefaultseppunct}\relax
		\EndOfBibitem
		\bibitem[Elser(2003)]{Elser03:JOSAA20}
		Elser,~V. Phase Retrieval by Iterated Projections. \emph{J. Opt. Soc. Am. A}
		\textbf{2003}, \emph{20}, 40--55\relax
		\mciteBstWouldAddEndPuncttrue
		\mciteSetBstMidEndSepPunct{\mcitedefaultmidpunct}
		{\mcitedefaultendpunct}{\mcitedefaultseppunct}\relax
		\EndOfBibitem
		\bibitem[Marchesini(2007)]{Marchesini07:RSI78}
		Marchesini,~S. Invited Article: A Unified Evaluation of Iterative Projection
		Algorithms for Phase Retrieval. \emph{Rev. Sci. Instrum.} \textbf{2007},
		\emph{78}, 011301\relax
		\mciteBstWouldAddEndPuncttrue
		\mciteSetBstMidEndSepPunct{\mcitedefaultmidpunct}
		{\mcitedefaultendpunct}{\mcitedefaultseppunct}\relax
		\EndOfBibitem
		\bibitem[Loh \latin{et~al.}(2010)Loh, Eisebitt, Flewett, and
		Elser]{Loh10:PRE82}
		Loh,~N.-T.~D.; Eisebitt,~S.; Flewett,~S.; Elser,~V. Recovering Magnetization
		Distributions from Their Noisy Diffraction Data. \emph{Phys. Rev. E}
		\textbf{2010}, \emph{82}, 061128\relax
		\mciteBstWouldAddEndPuncttrue
		\mciteSetBstMidEndSepPunct{\mcitedefaultmidpunct}
		{\mcitedefaultendpunct}{\mcitedefaultseppunct}\relax
		\EndOfBibitem
		\bibitem[Turner \latin{et~al.}(2011)Turner, Huang, Krupin, Seu, Parks, Kevan,
		Lima, Kisslinger, McNulty, Gambino, Mangin, Roy, and
		Fischer]{Turner11:PRL107}
		Turner,~J.~J.; Huang,~X.; Krupin,~O.; Seu,~K.~A.; Parks,~D.; Kevan,~S.;
		Lima,~E.; Kisslinger,~K.; McNulty,~I.; Gambino,~R. \latin{et~al.}  X-Ray
		Diffraction Microscopy of Magnetic Structures. \emph{Phys. Rev. Lett.}
		\textbf{2011}, \emph{107}, 033904\relax
		\mciteBstWouldAddEndPuncttrue
		\mciteSetBstMidEndSepPunct{\mcitedefaultmidpunct}
		{\mcitedefaultendpunct}{\mcitedefaultseppunct}\relax
		\EndOfBibitem
		\bibitem[Tripathi \latin{et~al.}(2011)Tripathi, Mohanty, Dietze, Shpyrko,
		Shipton, Fullerton, Kim, and McNulty]{Ashish11:PNAS108}
		Tripathi,~A.; Mohanty,~J.; Dietze,~S.~H.; Shpyrko,~O.~G.; Shipton,~E.;
		Fullerton,~E.~E.; Kim,~S.~S.; McNulty,~I. Dichroic Coherent Diffractive
		Imaging. \emph{Proc. Nat. Acad. Sci. U.S.A.} \textbf{2011}, \emph{108},
		13393--13398\relax
		\mciteBstWouldAddEndPuncttrue
		\mciteSetBstMidEndSepPunct{\mcitedefaultmidpunct}
		{\mcitedefaultendpunct}{\mcitedefaultseppunct}\relax
		\EndOfBibitem
		\bibitem[Trammell(1962)]{Trammell62:PR126}
		Trammell,~G.~T. Elastic Scattering at Resonance from Bound Nuclei. \emph{Phys.
			Rev.} \textbf{1962}, \emph{126}, 1045--1054\relax
		\mciteBstWouldAddEndPuncttrue
		\mciteSetBstMidEndSepPunct{\mcitedefaultmidpunct}
		{\mcitedefaultendpunct}{\mcitedefaultseppunct}\relax
		\EndOfBibitem
		\bibitem[Hannon and Trammell(1969)Hannon, and Trammell]{Hannon69:PR186}
		Hannon,~J.~P.; Trammell,~G.~T. M\"ossbauer Diffraction. II. Dynamical Theory of
		M\"ossbauer Optics. \emph{Phys. Rev.} \textbf{1969}, \emph{186},
		306--325\relax
		\mciteBstWouldAddEndPuncttrue
		\mciteSetBstMidEndSepPunct{\mcitedefaultmidpunct}
		{\mcitedefaultendpunct}{\mcitedefaultseppunct}\relax
		\EndOfBibitem
		\bibitem[Hill and McMorrow(1996)Hill, and McMorrow]{Hill96:ACA52}
		Hill,~J.~P.; McMorrow,~D.~F. Resonant Exchange Scattering: Polarization
		Dependence and Correlation Function. \emph{Acta Cryst. A} \textbf{1996},
		\emph{52}, 236--244\relax
		\mciteBstWouldAddEndPuncttrue
		\mciteSetBstMidEndSepPunct{\mcitedefaultmidpunct}
		{\mcitedefaultendpunct}{\mcitedefaultseppunct}\relax
		\EndOfBibitem
		\bibitem[Haverkort \latin{et~al.}(2007)Haverkort, Tanaka, Tjeng, and
		Sawatzky]{Haverkort07:PRL99}
		Haverkort,~M.~W.; Tanaka,~A.; Tjeng,~L.~H.; Sawatzky,~G.~A. Nonresonant
		Inelastic X-Ray Scattering Involving Excitonic Excitations: The Examples of
		NiO and CoO. \emph{Phys. Rev. Lett.} \textbf{2007}, \emph{99}, 257401\relax
		\mciteBstWouldAddEndPuncttrue
		\mciteSetBstMidEndSepPunct{\mcitedefaultmidpunct}
		{\mcitedefaultendpunct}{\mcitedefaultseppunct}\relax
		\EndOfBibitem
		\bibitem[Sokaras \latin{et~al.}(2012)Sokaras, Nordlund, Weng, Mori, Velikov,
		Wenger, Garachtchenko, George, Borzenets, Johnson, Qian, Rabedeau, and
		Bergmann]{Sokaras12:RSI83}
		Sokaras,~D.; Nordlund,~D.; Weng,~T.-C.; Mori,~R.~A.; Velikov,~P.; Wenger,~D.;
		Garachtchenko,~A.; George,~M.; Borzenets,~V.; Johnson,~B. \latin{et~al.}  {A
			High Resolution and Large Solid Angle X-ray Raman Spectroscopy End-station at
			the Stanford Synchrotron Radiation Lightsource}. \emph{Rev. Sci. Instrum.}
		\textbf{2012}, \emph{83}, 043112\relax
		\mciteBstWouldAddEndPuncttrue
		\mciteSetBstMidEndSepPunct{\mcitedefaultmidpunct}
		{\mcitedefaultendpunct}{\mcitedefaultseppunct}\relax
		\EndOfBibitem
		\bibitem[Platzman and Tzoar(1970)Platzman, and Tzoar]{Platzman70:PRB2}
		Platzman,~P.~M.; Tzoar,~N. Magnetic Scattering of X Rays from Electrons in
		Molecules and Solids. \emph{Phys. Rev. B} \textbf{1970}, \emph{2},
		3556--3559\relax
		\mciteBstWouldAddEndPuncttrue
		\mciteSetBstMidEndSepPunct{\mcitedefaultmidpunct}
		{\mcitedefaultendpunct}{\mcitedefaultseppunct}\relax
		\EndOfBibitem
		\bibitem[Hiraoka \latin{et~al.}(2015)Hiraoka, Takahashi, Wu, Lai, Tsuei, and
		Huang]{Hiraoka15:PRB91}
		Hiraoka,~N.; Takahashi,~M.; Wu,~W.~B.; Lai,~C.~H.; Tsuei,~K.~D.; Huang,~D.~J.
		Magnetic Circular dichroism of Nonresonant X-ray Raman Scattering.
		\emph{Phys. Rev. B} \textbf{2015}, \emph{91}, 241112\relax
		\mciteBstWouldAddEndPuncttrue
		\mciteSetBstMidEndSepPunct{\mcitedefaultmidpunct}
		{\mcitedefaultendpunct}{\mcitedefaultseppunct}\relax
		\EndOfBibitem
		\bibitem[Waterfield~Price \latin{et~al.}(2016)Waterfield~Price, Johnson,
		Saenrang, Maccherozzi, Dhesi, Bombardi, Chmiel, Eom, and
		Radaelli]{Waterfield16:PRL117}
		Waterfield~Price,~N.; Johnson,~R.~D.; Saenrang,~W.; Maccherozzi,~F.;
		Dhesi,~S.~S.; Bombardi,~A.; Chmiel,~F.~P.; Eom,~C.-B.; Radaelli,~P.~G.
		Coherent Magnetoelastic Domains in Multiferroic $\mathrm{BiFeO}_{3}$ Films.
		\emph{Phys. Rev. Lett.} \textbf{2016}, \emph{117}, 177601\relax
		\mciteBstWouldAddEndPuncttrue
		\mciteSetBstMidEndSepPunct{\mcitedefaultmidpunct}
		{\mcitedefaultendpunct}{\mcitedefaultseppunct}\relax
		\EndOfBibitem
		\bibitem[Lutman \latin{et~al.}(2016)Lutman, MacArthur, Ilchen, Lindahl, Buck,
		Coffee, Dakovski, Dammann, Ding, D{\"u}rr, Glaser, Gr{\"u}nert, Hartmann,
		Hartmann, Higley, Hirsch, Levashov, Marinelli, Maxwell, Mitra, Moeller,
		Osipov, Peters, Planas, Shevchuk, Schlotter, Scholz, Seltmann, Viefhaus,
		Walter, Wolf, Huang, and Nuhn]{Lutman16}
		Lutman,~A.~A.; MacArthur,~J.~P.; Ilchen,~M.; Lindahl,~A.~O.; Buck,~J.;
		Coffee,~R.~N.; Dakovski,~G.~L.; Dammann,~L.; Ding,~Y.; D{\"u}rr,~H.~A.
		\latin{et~al.}  Polarization Control in an X-ray Free-electron Laser.
		\emph{Nat. Photon.} \textbf{2016}, \emph{10}, 468--472\relax
		\mciteBstWouldAddEndPuncttrue
		\mciteSetBstMidEndSepPunct{\mcitedefaultmidpunct}
		{\mcitedefaultendpunct}{\mcitedefaultseppunct}\relax
		\EndOfBibitem
		\bibitem[Lepard(1970)]{lepard70:CJP48}
		Lepard,~D.~W. Theoretical Calculations of Electronic Raman Effects of the
		$\mathrm{NO}$ and $\mathrm{O}_2$ Molecules. \emph{Can. J. Phys.}
		\textbf{1970}, \emph{48}, 1664--1674\relax
		\mciteBstWouldAddEndPuncttrue
		\mciteSetBstMidEndSepPunct{\mcitedefaultmidpunct}
		{\mcitedefaultendpunct}{\mcitedefaultseppunct}\relax
		\EndOfBibitem
		\bibitem[Blume and Gibbs(1988)Blume, and Gibbs]{Blume88:PRB37}
		Blume,~M.; Gibbs,~D. Polarization Dependence of Magnetic X-ray Scattering.
		\emph{Phys. Rev. B} \textbf{1988}, \emph{37}, 1779--1789\relax
		\mciteBstWouldAddEndPuncttrue
		\mciteSetBstMidEndSepPunct{\mcitedefaultmidpunct}
		{\mcitedefaultendpunct}{\mcitedefaultseppunct}\relax
		\EndOfBibitem
		\bibitem[Bartell and Gavin(1964)Bartell, and Gavin]{Bartell64:JACS86}
		Bartell,~L.~S.; Gavin,~R.~M. Effects of Electron Correlation in X-Ray and
		Electron Diffraction. \emph{J. Am. Chem. Soc.} \textbf{1964}, \emph{86},
		3493--3498\relax
		\mciteBstWouldAddEndPuncttrue
		\mciteSetBstMidEndSepPunct{\mcitedefaultmidpunct}
		{\mcitedefaultendpunct}{\mcitedefaultseppunct}\relax
		\EndOfBibitem
		\bibitem[Blume(1985)]{Blume85:JAP57}
		Blume,~M. Magnetic Scattering of X-rays. \emph{J. Appl. Phys.} \textbf{1985},
		\emph{57}, 3615--3618\relax
		\mciteBstWouldAddEndPuncttrue
		\mciteSetBstMidEndSepPunct{\mcitedefaultmidpunct}
		{\mcitedefaultendpunct}{\mcitedefaultseppunct}\relax
		\EndOfBibitem
		\bibitem[Schülke(2007)]{BookXrayScatter}
		Schülke,~W. \emph{Electron Dynamics by Inelastic X-Ray Scattering}; Oxford
		University Press: New York, 2007\relax
		\mciteBstWouldAddEndPuncttrue
		\mciteSetBstMidEndSepPunct{\mcitedefaultmidpunct}
		{\mcitedefaultendpunct}{\mcitedefaultseppunct}\relax
		\EndOfBibitem
		\bibitem[Malmqvist \latin{et~al.}(1990)Malmqvist, Rendell, and
		Roos]{Malmqvist90:JPC94}
		Malmqvist,~P.~{\AA}.; Rendell,~A.; Roos,~B.~O. The Restricted Active Space
		Self-consistent-field Method, Implemented with a Split Graph Unitary Group
		Approach. \emph{J. Phys. Chem.} \textbf{1990}, \emph{94}, 5477--5482\relax
		\mciteBstWouldAddEndPuncttrue
		\mciteSetBstMidEndSepPunct{\mcitedefaultmidpunct}
		{\mcitedefaultendpunct}{\mcitedefaultseppunct}\relax
		\EndOfBibitem
		\bibitem[Emma \latin{et~al.}(2010)Emma, Akre, Arthur, Bionta, Bostedt, Bozek,
		Brachmann, Bucksbaum, Coffee, Decker, Ding, Dowell, Edstrom, Fisher, Frisch,
		Gilevich, Hastings, Hays, Hering, Huang, Iverson, Loos, Messerschmidt,
		Miahnahri, Moeller, Nuhn, Pile, Ratner, Rzepiela, Schultz, Smith, Stefan,
		Tompkins, Turner, Welch, White, Wu, Yocky, and Galayda]{Emma10:NatPhoton4}
		Emma,~P.; Akre,~R.; Arthur,~J.; Bionta,~R.; Bostedt,~C.; Bozek,~J.;
		Brachmann,~A.; Bucksbaum,~P.; Coffee,~R.; Decker,~F.~J. \latin{et~al.}  First
		Lasing and Operation of an Angstrom-wavelength Free-electron Laser.
		\emph{Nat. Photon.} \textbf{2010}, \emph{4}, 641--647\relax
		\mciteBstWouldAddEndPuncttrue
		\mciteSetBstMidEndSepPunct{\mcitedefaultmidpunct}
		{\mcitedefaultendpunct}{\mcitedefaultseppunct}\relax
		\EndOfBibitem
		\bibitem[Lambert \latin{et~al.}(2008)Lambert, Hara, Garzella, Tanikawa, Labat,
		Carre, Kitamura, Shintake, Bougeard, Inoue, Tanaka, Salieres, Merdji, Chubar,
		Gobert, Tahara, and Couprie]{Lambert08:NatPhys4}
		Lambert,~G.; Hara,~T.; Garzella,~D.; Tanikawa,~T.; Labat,~M.; Carre,~B.;
		Kitamura,~H.; Shintake,~T.; Bougeard,~M.; Inoue,~S. \latin{et~al.}  Injection
		of Harmonics Generated in Gas in a Free-electron Laser Providing Intense and
		Coherent Extreme-ultraviolet Light. \emph{Nat. Phys.} \textbf{2008},
		\emph{4}, 296--300\relax
		\mciteBstWouldAddEndPuncttrue
		\mciteSetBstMidEndSepPunct{\mcitedefaultmidpunct}
		{\mcitedefaultendpunct}{\mcitedefaultseppunct}\relax
		\EndOfBibitem
		\bibitem[Huang \latin{et~al.}(2017)Huang, Ding, Feng, Hemsing, Huang,
		Krzywinski, Lutman, Marinelli, Maxwell, and Zhu]{Huang17:PRL119}
		Huang,~S.; Ding,~Y.; Feng,~Y.; Hemsing,~E.; Huang,~Z.; Krzywinski,~J.;
		Lutman,~A.~A.; Marinelli,~A.; Maxwell,~T.~J.; Zhu,~D. Generating Single-Spike
		Hard X-Ray Pulses with Nonlinear Bunch Compression in Free-Electron Lasers.
		\emph{Phys. Rev. Lett.} \textbf{2017}, \emph{119}, 154801\relax
		\mciteBstWouldAddEndPuncttrue
		\mciteSetBstMidEndSepPunct{\mcitedefaultmidpunct}
		{\mcitedefaultendpunct}{\mcitedefaultseppunct}\relax
		\EndOfBibitem
		\bibitem[Lutman \latin{et~al.}(2016)Lutman, MacArthur, Ilchen, Lindahl, Buck,
		Coffee, Dakovski, Dammann, Ding, D\"urr, Glaser, Gr\"unert, Hartmann,
		Hartmann, Higley, Hirsch, Levashov, Marinelli, Maxwell, Mitra, Moeller,
		Osipov, Peters, Planas, Shevchuk, Schlotter, Scholz, Seltmann, Viefhaus,
		Walter, Wolf, Huang, and Nuhn]{Lutman16:NatPhoton10}
		Lutman,~A.~A.; MacArthur,~J.~P.; Ilchen,~M.; Lindahl,~A.~O.; Buck,~J.;
		Coffee,~R.~N.; Dakovski,~G.~L.; Dammann,~L.; Ding,~Y.; D\"urr,~H.~A.
		\latin{et~al.}  Polarization Control in an X-ray Free-electron Laser.
		\emph{Nat. Photon.} \textbf{2016}, \emph{10}, 468--472\relax
		\mciteBstWouldAddEndPuncttrue
		\mciteSetBstMidEndSepPunct{\mcitedefaultmidpunct}
		{\mcitedefaultendpunct}{\mcitedefaultseppunct}\relax
		\EndOfBibitem
	\end{mcitethebibliography}
\end{document}


\maketitle
	
	\section{Analytical Derivation of Differential Scattering Cross Section of Nonresonant Magnetic X-ray Scattering (MXS)}
	In nonresonant MXS, the double differential scattering cross section $\frac{d^2\sigma}{d\Omega_2 d\hbar\omega_2}$ from initial state $\ket{i}$ to final state $\ket{f}$ is~\cite{Blume85:JAP57,BookXrayScatter}
	\begin{align}
		(\frac{d^2\sigma}{d\Omega_2 d\hbar\omega_2})_{\ket{i}\to\ket{f}}=&(\frac{e^2}{m_{\mathrm{e}}c^2})^2(\frac{\omega_2}{\omega_1})\lvert\bra{f}\sum_j e^{i(\bm{K}_1-\bm{K}_2)\cdot\bm{r}_j}\ket{i}(\bm{e}_1\cdot\bm{e}_2^*)-\frac{i\hbar}{m_{\mathrm{e}}c^2}\bra{f}\sum_je^{i(\bm{K}_1-\bm{K}_2)\cdot\bm{r}_j}\notag\\
		&[ic(\bm{p}_j\times\frac{\hat{\bm{K}}_1-\hat{\bm{K}_2}}{\hbar})\cdot\bm{P}_1+\bm{D}\cdot\frac{\bm{\sigma}_j}{2}]\ket{i}\rvert^2\delta(E_i-E_f+\hbar\omega_1-\hbar \omega_2),\label{eq:ddscs}
	\end{align}
	where $\omega_{1(2)}$, $\bm{K}_{1(2)}$ and $\bm{e}_{1(2)}$ are frequencies, wave vectors and polarization vectors of incident and outgoing x-ray. And $\bm{r}_{j}$, $\bm{p}_{j}$ and $\bm{\sigma}_{j}$ are position, momentum and spin operators of the $j$th electron. The vector functions $\bm{P}_1$ and $\bm{D}$ are expressed as
	\begin{align}
		\bm{P}_1=&\bm{e}_1\times\bm{e}_2^*\\
		\bm{D}=&\frac{\omega_1+\omega_2}{2}[\bm{e}_2^*\times\bm{e}_1-(\hat{\bm{K}}_2\times\bm{e}_2^*)\times(\hat{\bm{K}}_1\times\bm{e}_1)]-[\omega_1(\bm{e}_2^*\cdot\hat{\bm{K}}_1)(\hat{\bm{K}}_1\times\bm{e}_1)-\notag\\
		&\omega_2(\bm{e}_1\cdot
		\hat{\bm{K}}_2)(\hat{\bm{K}}_2\times\bm{e}_2^*)].
	\end{align}
	In nonresonant MXS, the photon energy of hard x-ray is much higher than the energy difference between molecular electronic states. 
	Thus it is reasonable to assume $\hbar\omega=\hbar\omega_1\approx\hbar\omega_2\gg|E_i-E_f|$. The integral of the $\delta$-function on the RHS of eq~\ref{eq:ddscs} results $\omega_2/\omega_1=1$, which also affects the form of $\bm{D}$.  
	Denote $\bm{q}=\bm{K}_1-\bm{K}_2$, one obtains
	\begin{align}
		(\frac{d\sigma}{d\Omega_2})_{\ket{i}\to\ket{f}}=&(\frac{e^2}{m_{\mathrm{e}}c^2})^2\lvert\bra{f}\sum_j e^{i\bm{q}\cdot\bm{r}_j}\ket{i}\rvert^2\lvert\bm{e}_1\cdot\bm{e}_2^*\rvert^2+(\frac{e^2}{m_{\mathrm{e}}c^2})^2\frac{2\hbar\omega}{m_{\mathrm{e}}c^2}\Re\{\bra{i}\sum_{j'}e^{-i\bm{q}\cdot\bm{r}_{j'}}[\frac{c}{\omega}(\bm{p}_{j'}\times\notag\\
		&\frac{\hat{\bm{K}}_1-\hat{\bm{K}_2}}{\hbar})\cdot\bm{P}_1^*+i\bm{P}_2^*\cdot\frac{\bm{\sigma}_{j'}}{2}]\ket{f}\bra{f}\sum_j e^{i\bm{q}\cdot\bm{r}_j}\ket{i}(\bm{e}_1\cdot\bm{e}_2^*)\},\label{eq:DCS_i_to_f}
	\end{align}
	where the square of the magnetic part is neglected since it is by $(\frac{\hbar\omega}{N_{\mathrm{e}}m_{\mathrm{e}}c^2})^2\approx 8\times10^{-5}$ smaller than the electric part (the first term in eq~\ref{eq:DCS_i_to_f}) and $\frac{\hbar\omega}{N_{\mathrm{e}}m_{\mathrm{e}}c^2}\approx 9\times10^{-3}$ smaller than the interference part (the second term in eq~\ref{eq:DCS_i_to_f}), where $N_{\mathrm{e}}$ is the number of electrons in a molecule, and 
	\begin{align}
		\bm{P}_2=\bm{e}_2^*\times\bm{e}_1-(\hat{\bm{K}}_2\times\bm{e}_2^*)\times(\hat{\bm{K}}_1\times\bm{e}_1)-(\bm{e}_2^*\cdot\hat{\bm{K}}_1)(\hat{\bm{K}}_1\times\bm{e}_1)+(\bm{e}_1\cdot
		\hat{\bm{K}}_2)(\hat{\bm{K}}_2\times\bm{e}_2^*).
	\end{align}

	Summing over all final states $\ket{f}$, we obtain the expression for the differential scattering cross section (DSCS) of non-resonant MXS as
	\begin{align}
		(\frac{d\sigma}{d\Omega})_{\text{tot}}=&(\frac{e^2}{m_{\mathrm{e}}c^2})^2\bra{i}\sum_{jj'} e^{i\bm{q}\cdot\bm{r}_{j,j'}}\ket{i}\lvert\bm{e}_1\cdot\bm{e}_2^*\rvert^2+(\frac{e^2}{m_{\mathrm{e}}c^2})^2\frac{2\hbar\omega}{m_{\mathrm{e}}c^2}\Re\bra{i}\sum_{jj'}e^{i\bm{q}\cdot\bm{r}_{j,j'}}\notag\\
		&[(\frac{c}{\omega}\bm{p}_{j'}\times\frac{\hat{\bm{K}}_1-\hat{\bm{K}_2}}{\hbar})\cdot\bm{P}_1^*+i\frac{\bm{\sigma}_{j'}}{2}\cdot\bm{P}_2^*](\bm{e}_1\cdot\bm{e}_2^*)\ket{i}.
	\end{align}
	
	\section{MXS Calculation with ab initio wave functions}
	In this section we present procedure to calculation the spin-dependent contribution to dichroic signal of MXS, using electronic wave functions of the molecule represented in the basis of configuration state functions (CSF), which are obtained from the restricted active space self-consistent field (RASSCF) method.
	The CSF $\ket{\Phi_k^{(S,M_S)}}$ can be written as a linear combination of different Slater determinants $\ket{\psi_{m}}$ composed of spin orbitals $\ket{\phi_{p}}$ of molecules.
	Thus the expectation of spin scattering operator $S=\sum_{jj'}e^{i\bm{q}\cdot\bm{r}_{j,j'}}i\frac{\bm{\sigma}_{j'}}{2}\cdot\Delta[\bm{P}_2^*(\bm{e}_1\cdot\bm{e}_2^*)]$ of state $\ket{\Psi}$ can be calculated as
	\begin{align}\label{eq:AB_MXS}
		\bra{\Psi}S(\bm{q})\ket{\Psi}=&\frac{i}{2}\Delta[\bm{P}_2^*(\bm{e}_1\cdot\bm{e}_2^*)]\cdot\sum_{k,l}\sum_{m,n}a_{k}^*a_{l}b_{k,m}^*b_{l,n}\{\bra{\psi_{m}}\sum_{j}\bm{\sigma}_{j}\ket{\psi_{n}}\notag\\
		&+\frac{1}{2}\bra{\psi_{m}}\sum_{j\neq j'}[e^{i\bm{q}\cdot\bm{r}_{j,j'}}\bm{\sigma}_{j'} + e^{i\bm{q}\cdot\bm{r}_{j',j}}\bm{\sigma}_{j}] \ket{\psi_{n}}\}\notag\\
		=&\frac{i}{2}\Delta[\bm{P}_2^*(\bm{e}_1\cdot\bm{e}_2^*)]\cdot[\int d\bm{r}_1 \sum_{p,q}\bm{\sigma}_{1,pq} \gamma_{pq}(\bm{r}_1)\notag\\
		&+\int d\bm{r}_1 d\bm{r}_2 \sum_{p,q,r,s}\left(e^{i\bm{q}\cdot\bm{r}_{12}}\bm{\sigma}_{2,qs} + e^{i\bm{q}\cdot\bm{r}_{21}}\bm{\sigma}_{1,pr}\right)\Gamma_{pqrs}(\bm{r}_1,\bm{r}_2)]
	\end{align}
	where $a_i$ and $b_{i,j}$ are linear combination coefficients respectively from $\ket{\Phi_i^{(S,M_S)}}$ to $\ket{\Psi}$ and from $\ket{\psi_{j}}$ to $\ket{\Phi_i^{(S,M_S)}}$, $\sigma_{i,pq}$ is the spin part of matrix elements $\bra{\phi_p(\bm{r}_i)}\bm{\sigma}\ket{\phi_q(\bm{r}_i)}$. $\gamma_{pq}(\bm{r}_1)$ and $\Gamma_{pqrs}(\bm{r}_1,\bm{r}_2)$ are one- and two-electron reduced density matrix elements, and 
	\begin{eqnarray}
		\sum_{pq}\gamma_{pq}(\bm{r}_1)&=&\rho(\bm{r}_1)\,, \\
		\sum_{pqrs}\Gamma_{pqrs}(\bm{r}_1,\bm{r}_2)&=&\rho^{(2)}(\bm{r}_1,\bm{r}_2)
		\,.
	\end{eqnarray}

	In eq~(\ref{eq:AB_MXS}), only $z$-component of $\bm{\sigma}_{i,pq}$ contributes to the final result because of the rotational averaging of molecular $x$ and $y$ spin axis. By denoting $\rho(\bm{r},\alpha(\beta))$ as spin-up(down) electron densities and $\rho_{\mathrm{s}}(\bm{r})=\rho(\bm{r},\alpha)-\rho(\bm{r},\beta)$ as spin density, the integral in eq~(\ref{eq:AB_MXS}) can be further expressed as
	\begin{align}
		&\int d\bm{r}_1 \sum_{p,q}\bm{\sigma}_{1,z,pq} \gamma_{pq}(\bm{r}_1)=\int d\bm{r}_1[\rho(\bm{r}_1,\alpha)-\rho(\bm{r}_1,\beta)]=\int d\bm{r} \rho_{\mathrm{s}}(\bm{r}),\label{eq:j=jp}\\
		&\int d\bm{r}_1 d\bm{r}_2 \sum_{p,q,r,s}e^{i\bm{q}\cdot\bm{r}_{12}}\sigma_{2,z,qs}\Gamma_{pqrs}(\bm{r}_1,\bm{r}_2)\notag\\
		=&\int d\bm{r}_1 d\bm{r}_2e^{i\bm{q}\cdot\bm{r}_{12}}[\rho^{(2)}(\bm{r}_1,\alpha;\bm{r}_2,\beta)+\rho^{(2)}(\bm{r}_1,\beta;\bm{r}_2,\alpha)-\rho^{(2)}(\bm{r}_1,\alpha;\bm{r}_2,\beta)-\rho^{(2)}(\bm{r}_1,\beta;\bm{r}_2,\beta)],\label{eq:jneqjp}\notag\\
		=&\frac{N_{\mathrm{e}}-1}{2N_{\mathrm{e}}}\int d\bm{r}_1 d\bm{r}_2  e^{i\bm{q}\cdot\bm{r}_{12}}[\rho(\bm{r}_1,\alpha)+\rho(\bm{r}_1,\beta)]
		[\rho(\bm{r}_2,\alpha)-\rho(\bm{r}_2,\beta)]\notag\\
		=&\frac{N_{\mathrm{e}}-1}{2N_{\mathrm{e}}}\int d\bm{r}_1 d\bm{r}_2
		e^{i\bm{q}\cdot\bm{r}_{12}} \rho(\bm{r}_1) \rho_{\mathrm{s}}(\bm{r}_2)\,,
	\end{align}
	where $\rho^{(2)}(\bm{r}_1;\bm{r}_2)=\frac{N_{\mathrm{e}}-1}{2N_{\mathrm{e}}}\rho(\bm{r}_1)\rho(\bm{r}_2)$ by neglecting the electronic correlation effects, and $N_{\mathrm{e}}$ is the number of electrons. This approximation is valid when elastic scattering plays a dominant role in the MXS signal.
	
	In the quantum beating of NO, eq~(\ref{eq:j=jp}) makes no contribution since the population of spin-up and spin-down electrons are equal. Thus, the spin density can be retrieved from eq~(\ref{eq:jneqjp}) by dividing the MXS dichroic signal by electron density of the inner shell electrons. 
	
	In the ab initio calculation, $\rho_{\mathrm{s}}(\bm{r})$ and $\rho^{(2)}(\bm{r}_1,\alpha(\beta);\bm{r}_2,\alpha(\beta))$ are expanded with Gaussian type orbitals (GTO). The integrals in eq~(\ref{eq:j=jp}) and (\ref{eq:jneqjp}) can be calculated by
	\begin{align}
		&\int dx(x-x_1)^{L_1}(x-x_2)^{L_2}e^{-\gamma_1(x-x_1)^2}e^{-\gamma_2(x-x_2)^2}=C\int dx(x-x_1)^{L_1}(x-x_2)^{L_2}e^{-\gamma(x-p)^2}\notag\\
		&\int dx(x-x_1)^{L_1}(x-x_2)^{L_2}e^{-\gamma_1(x-x_1)^2}e^{-\gamma_2(x-x_2)^2}e^{iq_xx}\notag\\
		=&e^{iq_xp-\frac{q_x^2}{4\gamma}}C\int dx(x-x_1)^{L_1}(x-x_2)^{L_2}e^{-\gamma(x-\frac{iq_x}{2\gamma}-p)^2},
	\end{align}
	where $C=e^{\frac{\gamma_1\gamma_2}{\gamma_1+\gamma_2}(x_1-x_2)^2}$, $\gamma=\gamma_1+\gamma_2$ and $p=\frac{\gamma_1x_1+\gamma_2x_2}{\gamma_1+\gamma_2}$.
	
	\section{Calculation of $\Delta[\bm{P}_{1(2)}^*(\bm{e}_1\cdot\bm{e}_2^*)]$}

	In this section we present the explicit expression of $\Delta[\bm{P}_{1(2)}^*(\bm{e}_1\cdot\bm{e}_2^*)]$.
	We first express $\bm{e}_{1(2),\mathrm{R(L)}}$ and $\bm{K}_{1(2)}$ in lab frame coordinates. Assuming that the incident x-ray wave vector is along the $Z$ axis of the lab frame, the unit vector of incident wave vector $\bm{K}_1$ and left (right) circular polarization $\bm{e}_{1,\mathrm{L(R)}}$ are
	\begin{align}
		\hat{\bm{K}}_1= \begin{pmatrix}
			0&0&1
		\end{pmatrix}^T\qquad
		\hat{\bm{e}}_{1,\mathrm{L}}= \frac{\sqrt{2}}{2}\begin{pmatrix}
			1&i&0
		\end{pmatrix}^T\qquad
		\hat{\bm{e}}_{1,\mathrm{R}}= \frac{\sqrt{2}}{2}\begin{pmatrix}
			1&-i&0
		\end{pmatrix}^T.
	\end{align}
	The wave vector of outgoing x-ray has a deviation from the incident one and can be expressed with polar angle $\theta$ and azimuth angle $\phi$ as
	\begin{align}
		&\bm{K}_2=\begin{pmatrix}
			K_{2,x}&K_{2,y}&K_{2,z}
		\end{pmatrix}^T\longrightarrow \hat{\bm{K}}_2=\begin{pmatrix}
			\sin\theta\cos\phi&\sin\theta\sin\phi&\cos\theta
		\end{pmatrix}^T\notag\\
		&\cos\theta = \frac{K_{2,z}}{\sqrt{K_{2,x}^2+K_{2,y}^2+K_{2,z}^2}}\qquad\sin\theta=\frac{\sqrt{K_{2,x}^2+K_{2,y}^2}}{\sqrt{K_{2,x}^2+K_{2,y}^2+K_{2,z}^2}}\notag\\
		&\cos\phi=\frac{K_{2,x}}{\sqrt{K_{2,x}^2+K_{2,y}^2}}\qquad\qquad\quad\sin\phi=\frac{K_{2,y}}{\sqrt{K_{2,x}^2+K_{2,y}^2}}.
	\end{align}
	Because the circular polarization of the x-ray remains unchanged in nonresonant MXS process, the polarization vector of outgoing x-ray $\bm{e}_{2,\mathrm{L(R)}}$ is
	\begin{align}
		\bm{e}_{2,\mathrm{L(R)}}=&R_z(\phi)R_y(\theta)\bm{e}_{1,\mathrm{L(R)}}\notag\\
		=&\begin{pmatrix}
			\cos\phi & -\sin\phi & 0\\
			\sin\phi & \cos\phi & 0\\
			0 & 0 & 1
		\end{pmatrix}\begin{pmatrix}
			\cos\theta & 0 & \sin\theta\\
			0 & 1 & 0\\
			-\sin\theta & 0 & \cos\theta
		\end{pmatrix}\begin{pmatrix}
			\frac{\sqrt{2}}{2} \\ \pm i\frac{\sqrt{2}}{2} \\ 0
		\end{pmatrix}\notag\\
		=&\frac{\sqrt{2}}{2}\begin{pmatrix}
			\cos\phi\cos\theta\mp i\sin\phi&\sin\phi\cos\theta\pm i\cos\phi&-\sin\theta
		\end{pmatrix}^T.
	\end{align}
	The corresponding ($\bm{e}_1\cdot\bm{e}_2^*$), $P_1^*$ and $P_2^*$ are
	\begin{align}
		&(\bm{e}_1\cdot\bm{e}_2^*)_{\mathrm{L(R)}}=\frac{1}{2}(1+\cos\theta) \\
		&P^*_{1,\mathrm{L(R)}}=\frac{1}{2}\begin{pmatrix}
			\pm i\sin\theta&\sin\theta&(\pm i\cos\phi+\sin\phi)(\cos\theta+1)
		\end{pmatrix}^T\\
		&P^*_{2,\mathrm{L(R)}}=\frac{1}{2}\sin\theta\begin{pmatrix}
			\pm i\\1\\0
		\end{pmatrix}+ \frac{1}{2}\sin\theta \begin{pmatrix}
			-\sin\phi\mp i\cos\theta\cos\phi\\ \cos\phi\mp i\cos\theta\sin\phi\\ \pm i\sin\theta
		\end{pmatrix}.                                                                     
	\end{align}
	Finally, we obtain the important functions for the MXS calculation $\Delta[\bm{P}_2^*(\bm{e}_1\cdot\bm{e}_2^*)]$ and $\Delta[\bm{P}_2^*(\bm{e}_1\cdot\bm{e}_2^*)]$ as
	\begin{align}
		&\Delta[\bm{P}_1^*(\bm{e}_1\cdot\bm{e}_2^*)]= \frac{i}{2}\begin{pmatrix}
			\sin\theta(\cos\theta+1)\cos\phi\\
			\sin\theta(\cos\theta+1)\sin\phi\\
			(1+\cos\theta)^2
		\end{pmatrix},\label{eq:deltap1e1e2}\\
		&\Delta[\bm{P}_2^*(\bm{e}_1\cdot\bm{e}_2^*)]= \frac{i}{2}\begin{pmatrix}
			\sin^3\theta\cos\phi\\
			\sin^3\theta\sin\phi\\
			\sin^2\theta(1+\cos\theta)
		\end{pmatrix},\label{eq:deltap2e1e2}
	\end{align}
	where $\omega\approx\omega_1\approx\omega_2$ is the frequency of the hard x-ray field.
	From eq~\ref{eq:deltap2e1e2}, it is obvious that $\Delta[\bm{P}_{2}^*(\bm{e}_1\cdot\bm{e}_2^*)]$ is purely imaginary, this fact will be used in the next section~\ref{sec:MXS_analysis}.
	
	\section{Analysis for the Vanishing Part of Interference Term in MXS of Nitric Oxide Molecule}\label{sec:MXS_analysis}
	In the electron dynamics of $\mathrm{NO}$, the electronic states are represented by spin orbitals $\ket{\alpha(\beta)}$ and real-valued spatial orbitals $\ket{\pi_{x(y)}}$.
	The spin part of differential scattering cross section (DSCS) of MXS circular dichroism is proportional to 
	\begin{align}
		\Re\bra{i}\sum_{jj'}e^{i\bm{q}\cdot\bm{r}_{j,j'}}i\frac{\bm{\sigma}_{j'}}{2}\cdot\Delta[\bm{P}_2^*(\bm{e}_1\cdot\bm{e}_2^*)]\ket{i}.\label{eq:spin_DCS}
	\end{align}
	Since $\ket{i}$ represents the electronic state, i.e., the superposition of $\ket{\Pi_{1/2}}$ and $\ket{\Pi_{3/2}}$, which is probed the MXS and conforms to Hund case (a), the total spin of electrons is quantized along the aligned molecular axis, $j$ and $j'$ are electron indices.
	This can be also understood by the magnetic field produced by orbital current of $\ket{\pi_{\pm}}$, which is a ring current around the molecular axis.
	The magnetic field leads to an energy splitting for different electron spin along the N-O axis, determining the $z$ axis for the aligned spin, which coincides with the molecular axis.
	The stimulated Raman process determines the $x$ and $y$ axes of spin in the molecular coordinate system, which is the same for different stimulated molecules without rotational average~\cite{Kraus13:PRL111}. 
	However, the spin of $\mathrm{NO}$ has no preference for any direction perpendicular to the aligned molecular axis. 
	Thus the spin operators $\sigma_x$ and $\sigma_y$ in eq~\ref{eq:spin_DCS} should be angularly averaged. 
	Suppose the relative angle between $X(Y)$ of the lab frame for $\Delta[\bm{P}_2^*(\bm{e}_1\cdot\bm{e}_2^*)]$ and $x(y)$ axis for $\sigma_{x(y)}$ is $\theta'$, then eq~\ref{eq:spin_DCS} should be multiplied by $\cos\theta'$. 
	It leads to a zero-valued spin part after averaging since $\int_0^{2\pi}\cos\theta'd\theta'=0$. 
	The zero-valued expectation of the $\sigma_x$ and $\sigma_y$ components in eq~\ref{eq:spin_DCS} is important since they correspond to scattering between different spin orbitals. 
	In other words, only scattering between the same spin orbital contributes to the final circular dichroism signal. 
	
	When the incident x-ray is parallel to the molecular axis that is aligned in the lab frame, the $z$ axis is the same as $Z$ axis.
	Since $\Delta[\bm{P}_{2}^*(\bm{e}_1\cdot\bm{e}_2^*)]$ is purely imaginary, the product between $i\sigma_z$ and $\Delta[\bm{P}_{2}^*(\bm{e}_1\cdot\bm{e}_2^*)]_{Z}$ is a real-valued function. eq~\ref{eq:spin_DCS} can be written as
	\begin{align}
		\Re\bra{i}\sum_{j<j'}(e^{i\bm{q}\cdot\bm{r}_{j,j'}}i\frac{\sigma_{z,j'}}{2}+e^{-i\bm{q}\cdot\bm{r}_{j,j'}}i\frac{\sigma_{z,j}}{2})\Delta[\bm{P}_2^*(\bm{e}_1\cdot\bm{e}_2^*)]_z\ket{i}\,.\label{eq:spin_DCS_1}
	\end{align} 
	and
	\begin{align}
		\Im\bra{i'}\sum_{j<j'}e^{i\bm{q}\cdot\bm{r}_{j,j'}}i\frac{\bm{\sigma}_{j'}}{2}\cdot\Delta[\bm{P}_2^*(\bm{e}_1\cdot\bm{e}_2^*)]\ket{i}=0\label{eq:0imag}
		\,.
	\end{align}
	where $\ket{i}$ and $\ket{i'}$ are the electronic states. 
	
	\section{Electron Dynamics and the Expression of DSCS in Terms of Density Matrix}
	
	With $\ket{\pi_{x(y)}}$ and $\ket{\alpha(\beta)}$, the electronic states $\ket{\Pi_{1/2(3/2)},+}$ are
	\begin{align}
		\ket{^{2}\Pi_{\frac{1}{2}},+}=&\frac{1}{\sqrt{2}}(\ket{\pi^+\beta}+\ket{\pi^-\alpha})=\frac{1}{2}[(\ket{\pi_x}+i\ket{\pi_y})\ket{\beta}+(\ket{\pi_x}-i\ket{\pi_y})\ket{\alpha}]\\
		\ket{^{2}\Pi_{\frac{3}{2}},+}=&\frac{1}{\sqrt{2}}(\ket{\pi^+\alpha}+\ket{\pi^-\beta})=\frac{1}{2}[(\ket{\pi_x}+i\ket{\pi_y})\ket{\alpha}+(\ket{\pi_x}-i\ket{\pi_y})\ket{\beta}].
	\end{align}
	The quantum beating dynamics for $\ket{\Psi^{+}(t)}=\frac{1}{\sqrt{2}}(\ket{^{2}\Pi_{\frac{1}{2}},+}+\ket{^{2}\Pi_{\frac{3}{2}},+}e^{-i\Delta Et/\hbar})$ is represented by
	\begin{align}
		\ket{\Psi^{+}(t)}&=\frac{1}{2\sqrt{2}}\ket{\alpha}[(1+e^{-i2\pi t/T})\ket{\pi_x}-i(1-e^{-i2\pi t/T})\ket{\pi_y}]\notag\\
		&+\frac{1}{2\sqrt{2}}\ket{\beta}[(1+e^{-i2\pi t/T})\ket{\pi_x}+i(1-e^{-i2\pi t/T})\ket{\pi_y}].
	\end{align}
	Thus the wave functions of the superposition state at $t=0, \frac{T}{4}, \frac{T}{2}, \frac{3T}{4}, T$ are
	\begin{align}
		&\ket{\Psi^{+}(0)}=\frac{1}{\sqrt{2}}(\ket{\alpha}+\ket{\beta})\ket{\pi_x}\notag\\
		&\ket{\Psi^{+}(T/4)}=\frac{1-i}{2\sqrt{2}}[\ket{\alpha}(\ket{\pi_x}+\ket{\pi_y})+\ket{\beta}(\ket{\pi_x}-\ket{\pi_y})]\notag\\
		&\ket{\Psi^{+}(T/2)}=\frac{-i}{\sqrt{2}}(\ket{\alpha}+\ket{\beta})\ket{\pi_x}\\
		&\ket{\Psi^{+}(3T/4)}=\frac{1+i}{2\sqrt{2}}[\ket{\alpha}(\ket{\pi_x}-\ket{\pi_y})+\ket{\beta}(\ket{\pi_x}+\ket{\pi_y})]\notag\\
		&\ket{\Psi^{+}(T)}=\frac{1}{\sqrt{2}}(\ket{\alpha}+\ket{\beta})\ket{\pi_x}.\notag
	\end{align}
	By defining $S_{\mathrm{R(I)}}=\Re(\Im)\{\sum_{jj'}e^{i\bm{q}\cdot\bm{r}_{j,j'}}i\frac{\bm{\sigma}_{j'}}{2}\cdot\Delta[\bm{P}_2^*(\bm{e}_1\cdot\bm{e}_2^*)]\}$, the four integrals corresponding to the elements in density matrix of two electronic states $\ket{^{2}\Pi_{\frac{1}{2}},+}$ and $\ket{^{2}\Pi_{\frac{3}{2}},+}$ are
	\begin{equation}
		\Re[\bra{ ^2{\Pi}_{\frac{1}{2}},+}S\ket{^2{\Pi}_{\frac{1}{2}},+}]=\frac{1}{4}\begin{pmatrix}
			\bra{\alpha\pi_x}\\
			\bra{\beta\pi_x}\\
			\bra{\alpha\pi_y}\\
			\bra{\beta\pi_y}
		\end{pmatrix}^{T}(S_{\mathrm{R}}\begin{pmatrix}
			1&1&0&0\\
			1&1&0&0\\
			0&0&1&-1\\
			0&0&-1&1
		\end{pmatrix}+S_{\mathrm{I}}\begin{pmatrix}
			0&0&1&-1\\
			0&0&1&-1\\
			-1&-1&0&0\\
			1&1&0&0
		\end{pmatrix})\begin{pmatrix}
			\ket{\alpha\pi_x}\\
			\ket{\beta\pi_x}\\
			\ket{\alpha\pi_y}\\
			\ket{\beta\pi_y}
		\end{pmatrix}\label{eq:rho11}
	\end{equation}
	\begin{equation}
		\Re[\bra{ ^2{\Pi}_{\frac{1}{2}},+}S\ket{^2{\Pi}_{\frac{3}{2}},+}]=\frac{1}{4}\begin{pmatrix}
			\bra{\alpha\pi_x}\\
			\bra{\beta\pi_x}\\
			\bra{\alpha\pi_y}\\
			\bra{\beta\pi_y}
		\end{pmatrix}^{T}(S_{\mathrm{R}}\begin{pmatrix}
			1&1&0&0\\
			1&1&0&0\\
			0&0&-1&1\\
			0&0&1&-1
		\end{pmatrix}+S_{\mathrm{I}}\begin{pmatrix}
			0&0&-1&1\\
			0&0&-1&1\\
			-1&-1&0&0\\
			1&1&0&0
		\end{pmatrix})\begin{pmatrix}
			\ket{\alpha\pi_x}\\
			\ket{\beta\pi_x}\\
			\ket{\alpha\pi_y}\\
			\ket{\beta\pi_y}
		\end{pmatrix}\label{eq:rho12}
	\end{equation}
	\begin{equation}
		\Re[\bra{ ^2{\Pi}_{\frac{3}{2}},+}S\ket{^2{\Pi}_{\frac{1}{2}},+}]=\frac{1}{4}\begin{pmatrix}
			\bra{\alpha\pi_x}\\
			\bra{\beta\pi_x}\\
			\bra{\alpha\pi_y}\\
			\bra{\beta\pi_y}
		\end{pmatrix}^{T}(S_{\mathrm{R}}\begin{pmatrix}
			1&1&0&0\\
			1&1&0&0\\
			0&0&-1&1\\
			0&0&1&-1
		\end{pmatrix}+S_{\mathrm{I}}\begin{pmatrix}
			0&0&1&-1\\
			0&0&1&-1\\
			1&1&0&0\\
			-1&-1&0&0
		\end{pmatrix})\begin{pmatrix}
			\ket{\alpha\pi_x}\\
			\ket{\beta\pi_x}\\
			\ket{\alpha\pi_y}\\
			\ket{\beta\pi_y}
		\end{pmatrix}\label{eq:rho21}
	\end{equation}
	\begin{equation}
		\Re[\bra{ ^2{\Pi}_{\frac{3}{2}},+}S\ket{^2{\Pi}_{\frac{3}{2}},+}]=\frac{1}{4}\begin{pmatrix}
			\bra{\alpha\pi_x}\\
			\bra{\beta\pi_x}\\
			\bra{\alpha\pi_y}\\
			\bra{\beta\pi_y}
		\end{pmatrix}^{T}(S_{\mathrm{R}}\begin{pmatrix}
			1&1&0&0\\
			1&1&0&0\\
			0&0&1&-1\\
			0&0&-1&1
		\end{pmatrix}+S_{\mathrm{I}}\begin{pmatrix}
			0&0&-1&1\\
			0&0&-1&1\\
			1&1&0&0\\
			-1&-1&0&0
		\end{pmatrix})\begin{pmatrix}
			\ket{\alpha\pi_x}\\
			\ket{\beta\pi_x}\\
			\ket{\alpha\pi_y}\\
			\ket{\beta\pi_y}
		\end{pmatrix},\label{eq:rho22}
	\end{equation}
	where the expectation of $S_{\mathrm{I}}$ is $0$ in light of the analysis in eq~\ref{eq:0imag}. 
	For the expectation of $S_{\mathrm{R}}$, the sum of diagonal elements in the $4\times4$ matrix is always $0$ since the result of spin-up orbital cancels out with the result of spin-down orbital.
	Further, the $4\times4$ matrix can be divided as four $2\times 2$ matrices. Each off-diagonal element in these $2\times2$ matrices is also $0$ since it corresponds to the term with $\sigma_{x}$ and $\sigma_y$. 
	eq~\ref{eq:rho11} and eq~\ref{eq:rho22} are diagonal terms in density matrix of $\ket{\Pi_{1/2(3/2)},+}$, which have no contribution to the circular dichroism. 
	For the off-diagonal terms eq~\ref{eq:rho12} and eq~\ref{eq:rho21} of the density matrix, the complex coefficient $c_{12}=a_{12}+ib_{12}$ in front of the $\bra{ ^2{\Pi}_{\frac{3}{2}},+}S\ket{^2{\Pi}_{\frac{1}{2}},+}$ gives an effective contribution to the DSCS, since 
	\begin{align}
		\Re[c_{12}\bra{ ^2{\Pi}_{\frac{3}{2}},+}S\ket{^2{\Pi}_{\frac{1}{2}},+}]=&\frac{1}{4}\begin{pmatrix}
			\bra{\alpha\pi_x}\\
			\bra{\beta\pi_x}\\
			\bra{\alpha\pi_y}\\
			\bra{\beta\pi_y}
		\end{pmatrix}^{T}S_{\mathrm{R}}\begin{pmatrix}
			a_{12}&a_{12}&b_{12}&-b_{12}\\
			a_{12}&a_{12}&b_{12}&-b_{12}\\
			b_{12}&b_{12}&-a_{12}&a_{12}\\
			-b_{12}&-b_{12}&a_{12}&-a_{12}
		\end{pmatrix}\begin{pmatrix}
			\ket{\alpha\pi_x}\\
			\ket{\beta\pi_x}\\
			\ket{\alpha\pi_y}\\
			\ket{\beta\pi_y}
		\end{pmatrix}\notag\\
		&+\frac{1}{4}\begin{pmatrix}
			\bra{\alpha\pi_x}\\
			\bra{\beta\pi_x}\\
			\bra{\alpha\pi_y}\\
			\bra{\beta\pi_y}
		\end{pmatrix}^{T}S_{\mathrm{I}}\begin{pmatrix}
			-b_{12}&-b_{12}&a_{12}&-a_{12}\\
			-b_{12}&-b_{12}&a_{12}&-a_{12}\\
			a_{12}&a_{12}&b_{12}&-b_{12}\\
			-a_{12}&-a_{12}&-b_{12}&b_{12}
		\end{pmatrix}\begin{pmatrix}
			\ket{\alpha\pi_x}\\
			\ket{\beta\pi_x}\\
			\ket{\alpha\pi_y}\\
			\ket{\beta\pi_y}
		\end{pmatrix}.\label{eq:c12rho12}
	\end{align}
	And in eq~\ref{eq:c12rho12}, $\bra{\alpha\pi_y}b_{12}S_{\mathrm{R}}\ket{\alpha\pi_x}-\bra{\beta\pi_y}b_{12}S_{\mathrm{R}}\ket{\beta\pi_x}+\bra{\alpha\pi_x}b_{12}S_{\mathrm{R}}\ket{\alpha\pi_y}-\bra{\beta\pi_x}b_{12}S_{\mathrm{R}}\ket{\beta\pi_y}=4\bra{\alpha\pi_y}b_{12}$ $S_{\mathrm{R}}\ket{\alpha\pi_x}$ . 
	For $c_{21}=c_{12}^*$ in front of the term $\bra{ ^2{\Pi}_{\frac{3}{2}},+}S\ket{^2{\Pi}_{\frac{1}{2}},+}$, the result is the same. 
	Finally, the result of circular dichroism is $2\times\frac{1}{4}\times4\bra{\alpha\pi_y}b_{12}S_{\mathrm{R}}\ket{\alpha\pi_x}=2\bra{\alpha\pi_y}b_{12}S_{\mathrm{R}}\ket{\alpha\pi_x}$.
	
	Previous discussion indicates a zero-valued signal in the circular dichroism before the stimulated Raman scattering excites $\mathrm{NO}$ to the electronic state $\ket{ ^2\Pi_{\frac{3}{2}},+}$. After $\mathrm{NO}$ molecules are excited to the superposition state, $b_{12}(t)$ can be retrieved from time-dependent circular dichroism pattern. 
	For the superposition state 
	\begin{align}
		\ket{\Psi^{+}(t)}=\frac{1}{\sqrt{1+|c_1|^2}}(\ket{^{2}\Pi_{\frac{1}{2}},+}+|c_1|e^{i\varphi}\ket{^{2}\Pi_{\frac{3}{2}},+}e^{-i\Delta Et/\hbar}),
	\end{align}
	and $b_{12}=\frac{|c_{1}|\sin(\omega_0 t-\varphi)}{1+|c_1|^2}$, where $\omega_0=\frac{\Delta E}{\hbar}$. The density matrix of the electronic states can be reconstructed from the time-dependent MXS since both $\varphi$ and $|c_1|$ can be retrieved with $b_{12}(t)$. For long time evolution, quantum decoherence leads to a decreasing $|\rho_{12}|$ ($\ket{1}$ and $\ket{2}$ are respectively $\ket{\Pi_{1/2},+}$ and $\ket{\Pi_{3/2},+}$), which results in the damped intensity of the magnetic x-ray scattering circular dichroism. 
	Following decoherence, the measured time-dependent $b_{12}(t)$ is a decaying function. 
	One can still extract the phase of the decaying function to retrieve $c_{12}$ from $b_{12}$, and the off-diagonal density matrix elements can be reconstructed.

	\section{Estimate of the Photon Number with XFEL}
	The ultrafast nonresonant MXS can be implemented by x-ray free electron laser (XFEL). 
	In this section, we demonstrate the feasibility to use MXS circular dichroism as a probe of coherent spin dynamics, we calculate the rate of scattered photon number with XFEL parameters. Within unit solid angle, the total rate of scattered photon number $N_{\mathrm{photon}}$ is 
	\begin{align}
		N_{\mathrm{photon}}=\frac{d\sigma}{d\Omega}\nu_{\mathrm{pulse}}N_{\mathrm{pulse}}N_{\mathrm{mol}},\label{eq:N_rate}
	\end{align}
	where $\nu_{\mathrm{pulse}}$ is the beam rate, $N_{\mathrm{pulse}}$ is the incident photon number for a single pulse and $N_{\mathrm{mol}}$ is the number of irradiated molecules per unit area at this moment. Further, $N_{\mathrm{mol}}$ can be expressed as $n_{\mathrm{mol}}L$. Here $n_{\mathrm{mol}}$ is molecular density and $L$ is the diameter of the molecular beam. Using practical experimental parameters of XFEL~\cite{Emma10:NatPhoton4,Barty13:ARPC64}, the photon number rate is estimated to be
	\begin{align}
		N_{\mathrm{photon}}=&\Delta(\frac{d\sigma}{d\Omega})\nu_{\mathrm{pulse}}N_{\mathrm{pulse}}n_{\mathrm{mol}}L\notag\\
		=&2\times10^{-5}~\mathrm{barn}\times10^6~\mathrm{Hz}\times10^{12}\times 10^{16}~\mathrm{cm^{-3}}\times0.1~\mathrm{cm}=2\times10^4~\mathrm{s}^{-1}.\label{eq:N_rate1}
	\end{align}
	Considering only $0.2\%$ of the NO molecules are typically electronically excited after stimulated Raman scattering, the final result of photon number rate is 
	\begin{align}
		N_{\mathrm{photon},0.2\%}= \frac{\sqrt{0.002}}{1+0.002} \frac{1+1}{\sqrt{1}}N_{\mathrm{photon}} \approx 1.8\times10^3~\mathrm{s^{-1}}
	\end{align}

	\providecommand{\latin}[1]{#1}
	\makeatletter
	\providecommand{\doi}
	{\begingroup\let\do\@makeother\dospecials
		\catcode`\{=1 \catcode`\}=2 \doi@aux}
	\providecommand{\doi@aux}[1]{\endgroup\texttt{#1}}
	\makeatother
	\providecommand*\mcitethebibliography{\thebibliography}
	\csname @ifundefined\endcsname{endmcitethebibliography}
	{\let\endmcitethebibliography\endthebibliography}{}